\documentclass[%
 reprint,
superscriptaddress,
 amsmath,amssymb,
 aps,
 prd
]{revtex4-2}
\usepackage{booktabs} 
\usepackage{graphicx}% Include figure files
\usepackage{dcolumn}% Align table columns on decimal point
\usepackage{bm}% bold math
\usepackage[mathlines]{lineno}% Enable numbering of text and display math
\usepackage[dvipsnames]{xcolor}
\usepackage[colorlinks=true, linkcolor=RoyalBlue, citecolor=ForestGreen, urlcolor=BrickRed]{hyperref}
\usepackage{physics}
\usepackage{orcidlink}
\newcommand{\sub}[1]{\ensuremath{_{\text{#1}}}}

\newcommand{\monedet}[0]{\ensuremath{m\sub{1,det}}}
\newcommand{\mtwodet}[0]{\ensuremath{m\sub{2,det}}}
\newcommand{\mone}[0]{\ensuremath{m\sub{1}}}
\newcommand{\mtwo}[0]{\ensuremath{m\sub{2}}}

\newcommand{\lumdist}[0]{\ensuremath{d\sub{L}}}
\newcommand{\reds}[0]{\ensuremath{z}}

\newcommand{\snr}{\ensuremath{\rm SNR}}

\usepackage{aasmacros}
\usepackage[normalem]{ulem}

\begin{document}

% \preprint{APS/123-QED}

\title{Quantifying the Scientific Potential of Intermediate and Extreme Mass Ratio Inspirals with the Laser Interferometer Space Antenna
% : Detection, Inference, and Instrument Degradation
% :\\with Forced Linebreak
}% Force line breaks with \\

\author{Lorenzo Speri
\orcidlink{0000-0002-5442-7267}}\email{lorenzo.speri@esa.int}
\affiliation{%
European Space Agency (ESA), European Space Research and Technology Centre (ESTEC), Keplerlaan 1, 2201 AZ Noordwijk, the Netherlands
}%%
\affiliation{Leiden Observatory, Leiden University, P.O. Box 9513, 2300 RA Leiden, the Netherlands}
\author{Francisco Duque\,\orcidlink{0000-0003-0743-6491
}}%
\affiliation{%
Max Planck Institute for Gravitational Physics (Albert Einstein Institute) Am Mühlenberg 1, 14476 Potsdam, Germany}
\author{Susanna~Barsanti\,\orcidlink{0000-0001-7321-2512}}
\affiliation{School of Mathematics and Statistics, University College Dublin, Belfield, Dublin 4, Ireland}
\author{Alessandro Santini\,\orcidlink{0000-0001-6936-8581}}
\affiliation{%
Max Planck Institute for Gravitational Physics (Albert Einstein Institute) Am Mühlenberg 1, 14476 Potsdam, Germany}
\author{Shubham Kejriwal\,\orcidlink{0009-0004-5838-1886}}
\affiliation{Department of Physics, National University of Singapore, Singapore 117551}
\author{Ollie~Burke\,\orcidlink{0000-0003-2393-209X}}
\affiliation{School of Physics and Astronomy, University of Glasgow, Glasgow G12 8QQ, UK}
\author{Christian E. A. Chapman-Bird\,\orcidlink{0000-0002-2728-9612}}
\affiliation{Institute for Gravitational Wave Astronomy \& School of Physics and Astronomy, University of Birmingham, Edgbaston, Birmingham B15 2TT, UK}

\date{\today}

\begin{abstract}
The Laser Interferometer Space Antenna (LISA) will enable precision studies of Extreme and Intermediate Mass Ratio Inspirals (EMRIs/IMRIs), providing unique probes of astrophysical environments of galactic nuclei and strong-field gravity. 
Using a fully relativistic pipeline across primary masses $m_1 \in [5\times10^4, 10^7]\,M_\odot$ and secondary masses $m_2 \in [1, 10^4]\,M_\odot$, we map instrumental performance directly to detection horizons and parameter measurement precision. EMRIs with $m_1 = 10^7\,M_\odot$ and $m_2 \sim 1\,M_\odot$ are the most sensitive to instrument degradation, with redshift horizons at $z \sim 0.01$, while IMRIs are the least sensitive to degradation and reach redshifts $z \sim 1-3$.
All prograde systems considered achieve sub-percent spin precision within three months of observation.
The full 4.5-year mission increases the horizon of systems with $m_1 = 10^7\,M_\odot$ and $m_2 \sim 1\,M_\odot$ by a factor of $\sim 4$ and improves sky localization by one to two orders of magnitude reaching $ < 10\,\mathrm{deg}^2$. 
IMRI detection is robust against degradation, but their parameter estimation is more vulnerable due to fewer cycles in band. 
With the full baseline, EMRI observations constrain scalar dipole emission and Kerr quadrupole deviations below ground-based bounds by one to two orders of magnitude.
We release the accompanying \href{https://github.com/lorenzsp/EMRI-FoM}{software} and an interactive \href{https://huggingface.co/spaces/lorenzsp/emri-imri-fom}{website} to enable the community to rapidly quantify the scientific potential of EMRIs and IMRIs.
\end{abstract}

\maketitle

% \tableofcontents

%%%%%%%%%%%%%%%%%%%%%%%%%%%%%%
\section{Introduction}
The Laser Interferometer Space Antenna (LISA) mission is a space-based gravitational wave (GW) observatory that was adopted by the European Space Agency (ESA) in January 2024 and is planned to be launched in the mid-2030s \cite{2024arXiv240207571C}.
The main scientific goal of LISA is to observe gravitational wave sources in the millihertz frequency range.
Key astrophysical sources in this frequency region are Extreme and Intermediate Mass Ratio Inspirals (EMRIs/IMRIs)~\cite{2019BAAS...51c..42B}. 
These systems consist of a stellar-mass compact object inspiralling into a massive black hole, producing long-lived gravitational wave signals. Such signals encode rich information for probing fundamental physics \cite{2013LRR....16....7G,2009PhRvD..80f4006S,2025arXiv250800516Z,2025PhRvD.112h4075Z,2026PhRvD.113b3036S,2022LRR....25....4A,2020PhLB..81135860P,2020GReGr..52...81B,2018MNRAS.478...28C,1997PhRvD..56.1845R,2007PhRvD..75d2003B,2011PhRvD..83j4027V,PhysRevD.102.124054,2024PhRvD.109j4079D, 2023PhRvL.131e1401B,2022PhRvD.106d4029B,2022NatAs...6..464M,2020PhRvL.125n1101M,2024PhRvD.109f4022S,2023PhRvD.107b3005Z}, astrophysical environments~\cite{2019CQGra..36n3001B,2023PhRvX..13b1035S,2011PhRvL.107q1103Y,2011PhRvD..84b4032K,2025PhRvD.111h4006D,2014PhRvD..89j4059B,2024PhRvD.110h4060K,2025PhRvD.111h2010K,2004CQGra..21S1595G,2025PhRvL.135u1401V}, black hole demographics \cite{2009CQGra..26i4034G,2023MNRAS.522.6043C,2026arXiv260115198S,2010PhRvD..81j4014G,2024arXiv241012408F}, and cosmology \cite{2021MNRAS.508.4512L,2025ApJ...991..223L}.
Their observations are expected to achieve the two key LISA mission science objectives identified in the LISA Definition Study Report \cite{2024arXiv240207571C}:
\emph{``Probe the properties and immediate environments of Black Holes in the local Universe''} and \emph{``Explore the fundamental nature of gravity and Black Holes''}.

Previous studies have extensively explored the scientific potential of these sources.
The most general study on EMRI science was performed in Ref.~\cite{2019BAAS...51c..42B}, where different astrophysical formation channels were considered to estimate detection rates and parameter estimation capabilities.
Other studies have linked the scientific achievement of the EMRI and IMRI objectives to specific mission performance requirements, such as mission duration in Ref.~\cite{2022GReGr..54....3A}, failure of laser measurements \cite{2008CQGra..25f5005V}, and instrument calibration in Ref.~\cite{2022PhRvD.106b2003S}.
This work is part of the ongoing effort for the definition of LISA Figures of Merit to assess the achievement of science objectives for all LISA sources, including EMRIs/IMRIs, to quantify the overall scientific performance of LISA~\cite{FoM_inprep}.

Valid performance requirements for the mission must be robust against the large uncertainties in astrophysical population rates, which span orders of magnitude~\cite{2017PhRvD..95j3012B,2021PhRvD.104f3007P}.
While the scientific value of these sources is clear, a systematic quantification of how instrumental degradation (sensitivity loss) and mission duration explicitly degrade the \emph{quality} of the science return (detection horizon and measurement precision) across the full parameter space is lacking.
Bridging this gap is essential for defining mission-critical thresholds that ensure science objectives are met regardless of the specific event rate.

In this work, we quantify the scientific potential of EMRIs and IMRIs by mapping LISA's instrumental performance directly to observable metrics.
Instead of relying on specific population models, we adopt a grid-based approach covering the relevant parameter space with primary masses $\mone \in [5\times 10^4, 10^7] M_{\odot}$ and secondary masses $\mtwo \in [1, 10^4] M_{\odot}$.
We utilize a fully relativistic waveform generation pipeline to compute Signal-to-Noise Ratios (SNR) and Fisher Information Matrices, assessing detection horizons and parameter estimation precision.
We formulate a ``degradation framework'' to translate these findings into illustrative instrumental performance criteria.

We find that the science reach for EMRIs and IMRIs responds differently to instrumental performance.
EMRIs with $m_1 \sim 10^7\,M_\odot$ and $m_2 \sim 1\,M_\odot$ are the most demanding sources: their detection horizon is most vulnerable to sensitivity degradation, and their sky localization improves by orders of magnitude with a nominal 4.5-year mission duration.
Conversely, IMRIs are robust in detection but their characterization is highly sensitive to noise levels due to their rapid evolution.
Crucially, we show that measuring the primary black hole spin with precision better than $10^{-3}$ is achievable even in conservative scenarios, while probing environmental effects requires the full mission duration.

We do not consider the science objectives associated with cosmology, as these require population models and are left for future work \cite{2021MNRAS.508.4512L}.

This work is organized as follows.
In Sections~\ref{subsec:science_objectives}--\ref{subsec:source_grid} we discuss how to convert the science objectives for EMRIs and IMRIs into a source grid of representative sources.
In Section~\ref{subsec:instrument_constraint} we discuss the LISA sensitivity, resolution, and operational lifetime.
In Section~\ref{subsec:pipeline} we define the data analysis quantities used in this work and how they are computed.
In Section~\ref{sec:results} we discuss how example instrumental performance criteria can be defined and present the results for the sources in the grid and which ones are going to be most affected.

%%%%%%%%%%%%%%%%%%%%%%%%%%%%%%
\section{Methods}
%%%%%%%%%%%%%%%%%%%%%%%%%%%%%%
\subsection{Science Objectives and Investigations}\label{subsec:science_objectives}
EMRIs and IMRIs are expected to probe the properties and immediate environments of black holes in the local Universe and explore the fundamental nature of gravity and black holes \cite{2024arXiv240207571C}.
Associated with these science objectives are Science Investigations (SIs) which for completeness we repeat below following Ref.~\cite{2024arXiv240207571C}.

SI 3.1 aims to detect gravitational-wave signals from EMRIs with massive black hole masses, $m_1$, between a few times $10^4 M_{\odot}$ and a few times $10^7 M_{\odot}$ at redshifts up to $\reds \sim 3$, for inspiralling objects of $m_2 \approx 10 M_{\odot}$. This investigation aims to address the following questions: What are the mass and spin distributions of massive black holes? In what types of stellar and gaseous environments do these massive black holes live? Which physical processes dominate stellar dynamics near these massive black holes?

SI 3.2 aims to detect gravitational waves from IMRIs at low redshift, $\reds < 2$, in which the intermediate-mass black hole has mass in the range $10^3-10^4 M_{\odot}$. How readily do intermediate mass black holes form in stellar clusters and Galactic nuclei? How do these intermediate-mass black holes subsequently grow and what are their properties? How often do these intermediate-mass black holes merge with massive black holes?

SI 5.2-5.5 aim to probe theories of gravity alternative to GR, for example: are the massive objects observed at the centres of galaxies consistent with the rotating Kerr solution predicted by GR, and
are there new fundamental fields that induce hair on black holes?~\cite{2017arXiv170200786A}.

{Finally, for SI 5.3, we probe various GW emission channels. For example, are there GW emission channels beyond GR? Are there new physical degrees of freedom and extra GW polarisations, as predicted by some extensions of the standard model and of GR?}

\begin{figure}[h!]
    \centering
    \includegraphics[width=0.99\columnwidth]{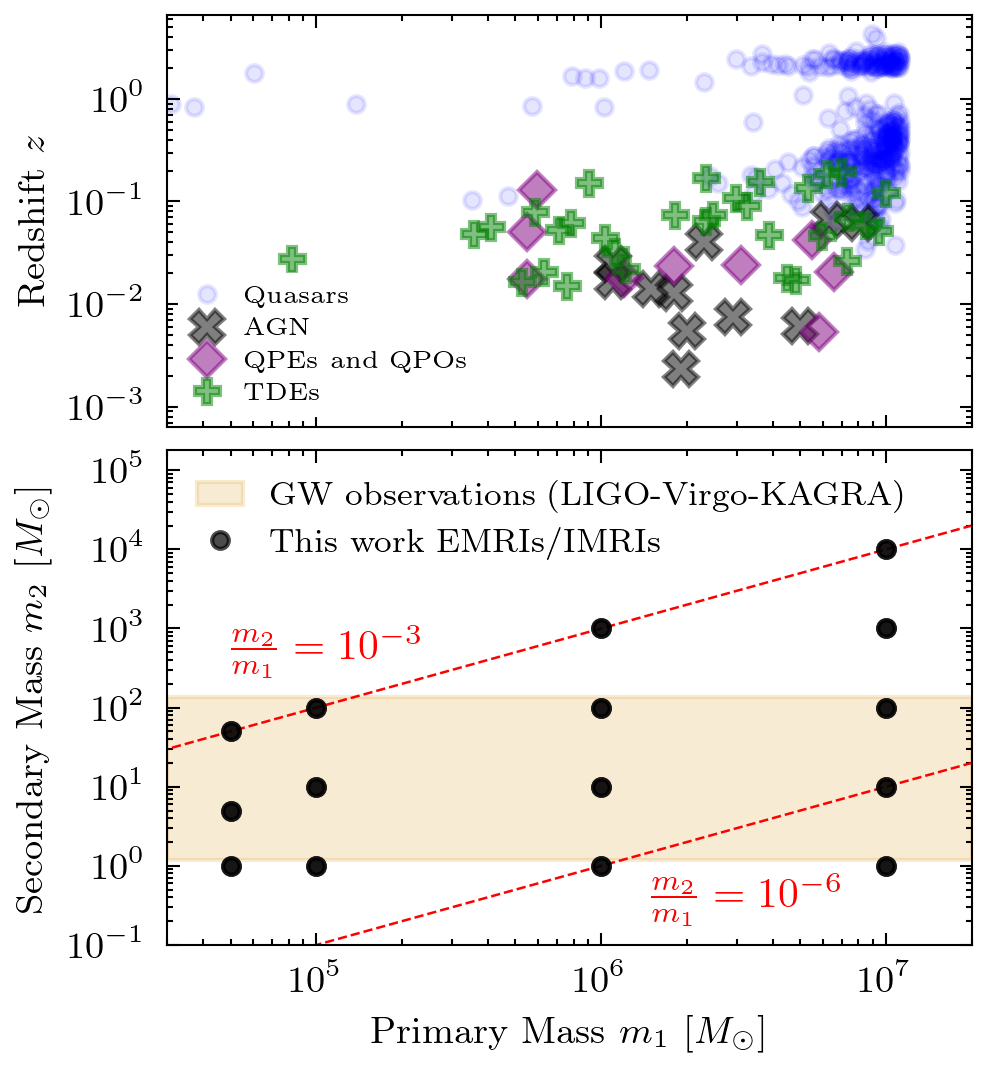}
    \caption{
    \textit{Top}: Electromagnetic observations of massive black holes with primary mass $m_1$ and redshift $z$ that could potentially host companions and therefore be the target of LISA observations.
    We present estimates of Quasars (blue dots), Active Galactic Nuclei (AGN) (black cross), Quasi Periodic Eruptions and Oscillations (QPE and QPO) (violet diamond), and Tidal Disruption Events (TDEs) (green plus).
    These redshift and mass measurements are characterized by large uncertainties and are only representative of the capabilities of electromagnetic observations.
    \textit{Bottom}: EMRIs and IMRIs component masses considered in this work, source-frame primary $m_1$ and secondary mass $m_2$.
    The horizontal shaded area (yellow) shows the ground-based masses from gravitational wave observations of the LIGO-Virgo-KAGRA collaboration.
    The diagonal lines indicate constant mass ratios $m_2/m_1$.
    }
    \label{fig:emri_imri_population}
\end{figure}

Given the uncertainties on the astrophysical population of EMRIs and IMRIs (detection rates range between 1-1000~\cite{2017PhRvD..95j3012B,2021PhRvD.104f3007P}),
we do not make an explicit assumption on the underlying formation channels and theoretical astrophysical distribution of these sources \cite{2025arXiv250900469S}.
Instead, we adopt the following principles for constructing a grid of the parameter space of interest for the science investigations:
\begin{itemize}
    \item we consider mass ranges within the validity of our waveform generation framework, here adopted to be FastEMRIWaveform (FEW), for eccentric equatorial inspirals into rapidly-spinning black holes implemented through the class \texttt{FastKerrEccentricEquatorialFlux} of FEW~\cite{2025PhRvD.112j4023C,2023arXiv230712585S,2021PhRvD.104f4047K,2021PhRvL.126e1102C,chapman_bird_2025_15630565};
    \item we ensure the secondary stellar-mass object mass in EMRIs spans the current range of masses observed by the LIGO-Virgo-KAGRA collaboration \cite{2025arXiv250818079T}; 
    \item we choose sources to span the LISA sensitivity band in order to cover the time-frequency evolution of EMRIs and IMRIs. 
\end{itemize}

\subsection{Source Assumptions}\label{subsec:source_grid}
The science objectives and investigations are translated into the source population presented in Fig.~\ref{fig:emri_imri_population}.
We consider sources with primary masses in the range
$\mone \in [5\times 10^4, 10^7] M_{\odot}$ 
and secondary masses in the range 
$\mtwo \in [1, 10^4] M_{\odot}$, 
with mass ratios $\mtwo/\mone$ spanning from $10^{-3}$ to $10^{-7}$. 

The primary masses cover the expected mass range of massive black holes observed with electromagnetic observations such as those from 
Sloan Digital Sky Survey Data Release 16 Quasar catalog (SDSS DR16Q)~\cite{2022ApJS..263...42W},
Active Galactic Nuclei 
(Table III of \cite{2025arXiv250103252L}), 
Quasi Periodic Oscillations and Eruptions (see Table 1 in Ref.~\citep{2024MNRAS.532.2143K}; see also Refs.~\citep{2019Natur.573..381M,2021Natur.592..704A,2024A&A...684A..64A,2020sea..confE..40G,2008arXiv0807.1899G,2013ApJ...776L..10L,2019Sci...363..531P,2024SciA...10J8898P,2024NatAs...8..347G}), and Tidal Disruption Events (TDEs) \cite{2024MNRAS.527.2452M}.
For more information on the electromagnetic observations see Appendix~\ref{app:em_observations}.
Smaller primary masses than these would lead to a GW frequency plunge above the LISA sensitivity band, while larger primary masses would lead to initial frequencies below it.
The secondary masses cover the expected mass range of stellar-mass black holes observed with ground-based detectors \cite{2025arXiv250818079T}.

Having defined the component masses, we then need to consider the orbits of the inspiralling object around the massive black hole.
We consider prograde and retrograde orbits labelled by the positive and negative sign of the dimensionless Kerr spin parameter $a$ of the primary massive black hole.
We neglect the spin of the secondary object, as its effect on the waveform is negligible at first order in the mass ratio expansion \cite{2020PhRvD.102b4041P,2024PhRvD.109l4048B} and currently not implemented.

\begin{figure*}
    \centering
    \includegraphics{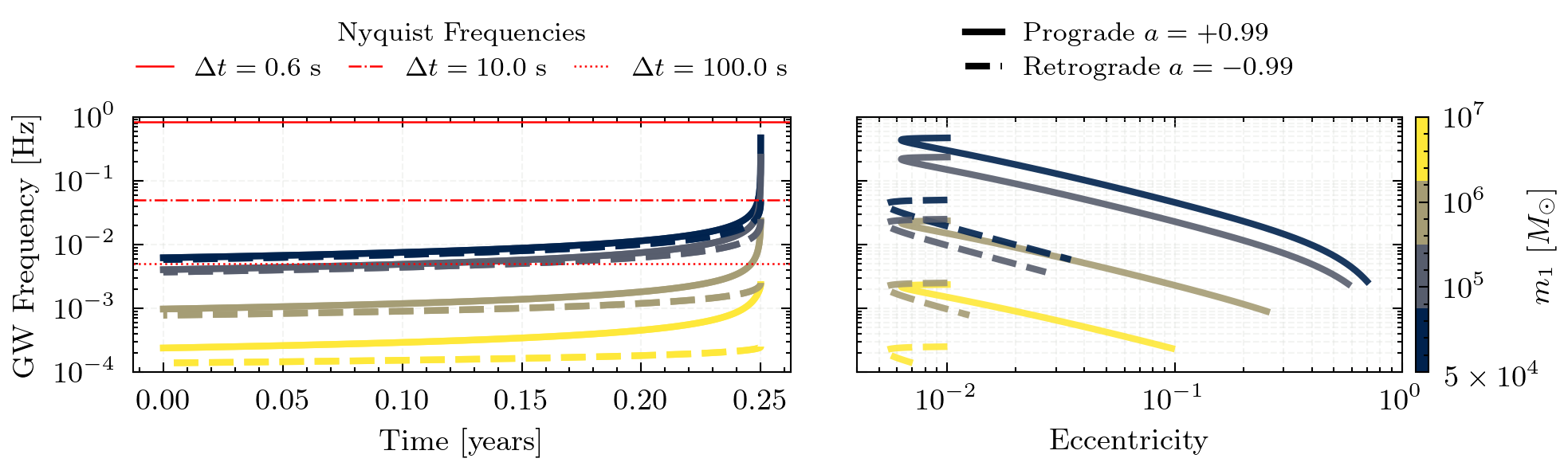}
    \caption{
    GW frequency evolution as a function of time (left) and eccentricity (right) for an inspiral of three months for different primary mass $m_1$ and prograde (solid line) and retrograde dimensionless spin $a$ (dashed line).
    The GW frequency is given by twice the azimuthal fundamental frequency of the orbit.
    The mass ratio is fixed to $\mtwo/\mone = 10^{-3}$ and the horizontal lines indicate the Nyquist frequency corresponding to different sampling rates $\Delta t$.
    }
    \label{fig:ecc_gwf_population}
\end{figure*}

Prograde orbits are more likely to be observed due to the higher frequency content of the signal and the longer time spent in band.
Retrograde orbits allow us to set conservative requirements on the instrumental capabilities due to their lower frequency content.
This difference is shown in the left panel of Figure~\ref{fig:ecc_gwf_population}, where the GW frequency evolution as a function of time is shown for different primary masses and at fixed mass ratio $\mtwo/\mone = 10^{-3}$.
We choose a reference absolute spin of $|a| = 0.99$ \cite{2013ApJ...762...68D}. This is below the Thorne limit $a \sim 0.998$ \cite{1974ApJ...191..507T} and above the median spin $\sim$ 0.98 considered in Ref.~\cite{2017PhRvD..95j3012B}, and is consistent with observations presented in Ref.~\cite{2021ARA&A..59..117R}.
We do not aim to model astrophysically motivated spin distributions, but rather to cover the parameter space of interest for the science investigations, from pessimistic to optimistic scenarios.

Although there are generic EMRI models \cite{2021PhRvD.104f4047K}, they have been shown to be inaccurate in estimating the SNR and the measurement precision for high spin and eccentricity \cite{2025PhRvD.112j4023C}.
Therefore, we consider equatorial orbits and we do not include orbital inclination as a parameter of the source population.

We consider two final eccentricities $e_f = 0, 10^{-2}$ to cover circular and eccentric orbits. The precise definition of final eccentricity will be given in Section~\ref{subsubsec:trajectory}.
Most of our analysis will be focused on circular orbits which are going to be more relevant for detection purposes, since, for fixed inspiral duration, larger eccentricities are likely to provide lower SNRs.
However, future investigations should include the impact of eccentricity and its relation to specific formation scenarios \cite{2025arXiv250900469S}, such as the loss cone \cite{2025arXiv250902394M}, wet EMRIs \cite{2025arXiv250103252L,2023MNRAS.521.4522D}, and tidal-separation \cite{2012MNRAS.425.2401A} formation scenarios.
Depending on the mass ratio and primary mass, the initial eccentricity can vary from nearly circular to highly eccentric \cite{2017PhRvD..95j3012B}.
This effect can be seen in the right panel of Fig.~\ref{fig:ecc_gwf_population} where the eccentricity evolution as a function of GW frequency is shown for different primary masses and at fixed mass ratio $\mtwo/\mone = 10^{-3}$.

To assess EMRIs' observations to probe astrophysical environments and modifications to GR, we use the 
parametrized post-Einsteinian framework~\cite{2009PhRvD..80l2003Y} which allows us to capture a wide range of physical effects in a model-agnostic way.
Since such a formalism cannot be easily generalized to generic orbits (although there are extensions to precessing orbits and bursts in Ref.~\cite{2014PhRvD..90j4010L,2023PhRvD.107d4046L}), we consider only circular EMRIs.
These effects are implemented through power-law corrections to the rescaled total angular momentum flux~\cite{2023PhRvX..13b1035S}:
\begin{equation}\label{eq:modifiedtorque}
\dot{L}/(\nu^2 m_1)
=
\dot{L}_{\rm GW}
+
A \left(\frac{p}{10 m_1}\right)^{n_r}
\cdot
\frac{32}{5}
\left( \frac{p}{m_1} \right)^{-7/2} \, ,
\end{equation}
where $\nu = m_1 m_2 /(m_1+m_2)^2$ is the symmetric mass ratio, the first term $\dot{L}_{\rm GW}$ is the gravitational-wave contribution and the second term incorporates the deviation from (vacuum) GR.
The dimensionless amplitude $A$ is the
fractional deviation at semi-latus rectum $p = 10m_1$ in the flux, while $n_r$ is a power-law index determined by the physical mechanism responsible for the departure from vacuum GR.

Different choices of $n_r$ describe the leading (weak-field) behaviors of both GR modifications and astrophysical environmental effects incorporating different science investigations:
\begin{itemize}
    \item $n_r = 8$ corresponds to migration torques in the fiducial $\alpha$-disk model for accretion-disk effects~\cite{2011PhRvD..84b4032K, 2011PhRvL.107q1103Y, 2014PhRvD..89j4059B};
    \item $n_r = 5.5$ corresponds to dynamical friction from a constant density dark matter distribution~\cite{Eda:2014kra, 2013ApJ...774...48M,  Kavanagh:2020cfn};
    \item $n_r = 1$ corresponds to dipolar emission of the scalar charge carried by the stellar-mass secondary compact object in a broad class of modified gravity theories~\cite{2024PhRvD.109j4079D,2022PhRvD.106d4029B,2022NatAs...6..464M,2020PhRvL.125n1101M,2026PhRvD.113b3036S,Gliorio:2026yvh,2024PhRvD.109f4022S};
    \item $n_r = -2$ corresponds to deviations from the Kerr multipole structure of massive black holes (i.e., violations of the no-hair theorem)~\cite{1995PhRvD..52.5707R, 2006CQGra..23.4167G, 2007PhRvD..75d2003B}.
\end{itemize}
For our analysis we assume no deviation and study the constraint on the amplitude $A$.
In the Appendix~\ref{appendix:beyond_GR}, we provide further details on these deviations and describe how the fractional amplitude $A$ can be mapped to physical quantities of interest (e.g., disk density and accretion rate for $n_r = 8$).
For the scalar charge emission $(n_r=1)$, we also consider the fully relativistic case, described in Appendix~\ref{appendix:scalar_charge}.

%%%%%%%%%%%%%%%%%%%%%%%%%%%%%%
\subsection{Instrumental constraints}\label{subsec:instrument_constraint}
%%%%%%%%%%%%%%%%%%%%%%%%%%%%%%

The LISA mission has temporal and operational constraints that affect its scientific capabilities \cite{2024arXiv240207571C}. 
In this section, we summarize the key instrumental constraints relevant to the purpose of this work.

The nominal science phase is scheduled to last $4.5$ years from the beginning of operational observations. 
However, this nominal period does not represent continuous uninterrupted observations. 
The mission availability, also referred to as the duty cycle, is specified as a percentage of the actual usable data assumed to be $82\%$. 
Recent work~\cite{burke_tdi_gaps} has demonstrated that the (median) effective duty cycle after accounting for the merging of data gaps is $\sim 84$\%, implying that our results (when assuming a duty cycle of $82\%$) should be interpreted as conservative. 
The duty cycle of the LISA mission accounts for scheduled maintenance intervals, instrument calibration activities, spacecraft decommissioning procedures, and unforeseen operational interruptions. 
When these factors are accounted for as data gaps, the actual accumulated science observing time is approximately $3.69$ years, which represents the effective duration of gravitational wave data suitable for scientific analysis.

In this work, we do not explicitly model the effect of data gaps, but this effect could be approximated by applying a factor of $\sqrt{0.82}$ to SNRs and $\sqrt{1/0.82}$ to precision statements.
% {Multiplying precision measurements of EMRI parameters by $\sim 0.82$ would be a reasonable approximation when accounting for data gaps due to the quasi-monochromatic features of the EMRI waveforms observed during the early inspiral. 
This approximation would start to break down near plunge, but would still provide a rough estimate to the loss of parameter precision and SNR.

The strain sensitivity is related to the instrument sensitivity through the transfer function of the detector, which accounts for the response of the interferometric arms to GW signals.
The instrument sensitivity is characterized by the noise power spectral density (PSD) of the Time-Delay Interferometry (TDI) data stream channels~\cite{2021arXiv210801167B}.

The LISA data processing framework employs TDI as its foundation to remove laser frequency noise while retaining GW signals~\cite{armstrong_1999_tdi, alg_approach_TDI, 2004PhRvD..69h2001T, geom_TDI_vallisneri}, with supporting techniques to suppress clock noise~\cite{2018PhRvD..98d2003T, clock...JB...olaf}, spacecraft jitter~\cite{2022JOpt...24f5601H, TTL1, TTL2, TTL3}, and other instrumental effects~\cite{RangingSensorFusion, 2019PhRvD..99h4023B}. 
We will express a degradation $d$ of the LISA sensitivity as an increase in the noise PSD of the TDI channels.

The fundamental performance limits are due to test mass acceleration noise at frequencies below approximately 1 mHz and photon shot noise at higher frequencies, with the former limitation validated through the LISA Pathfinder mission~\cite{2024PhRvD.110d2004A}. 
A practical constraint from the Galactic binary foreground exists in the frequency band from 0.3 to 3 mHz, but this represents astrophysical signals rather than instrumental limitations. 
\begin{figure}[h!]
    \centering
    \includegraphics[width=0.99\columnwidth]{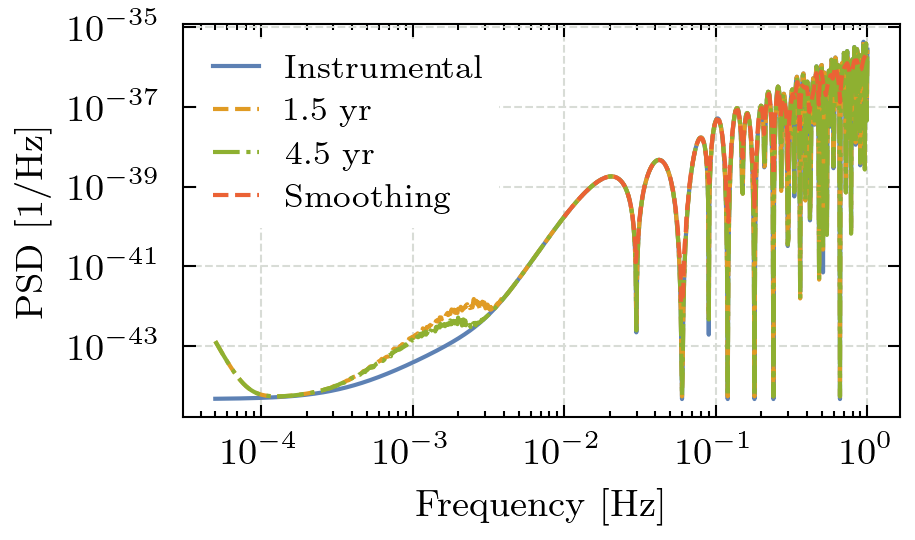}    
    \caption{LISA Time Delay Interferometry sensitivity curves for 2nd-generation A and E channels. 
    The blue solid line represents the instrumental noise only, while the dashed and dashed-dotted lines include the galactic white dwarf confusion foreground estimated for durations of 1.5 and 4.5 years, respectively.
    At frequencies above $10^{-2}$ Hz, we apply a smoothing of the sensitivity curve, shown as a dashed red line, to avoid numerical artifacts due to the zeros of the response.
    At frequencies below $10^{-4}$ Hz, we impose a smooth increase to account for the lower-frequency limit guaranteed by the requirements.
    }    
    \label{fig:psd_curves}
\end{figure}
To account for these instrumental and processing constraints, we utilize the LISA sensitivity curves specified for the TDI second-generation A and E channels shown in Fig.~\ref{fig:psd_curves}, which incorporate both instrumental noise according to Ref.~\cite{2021arXiv210801167B} and the galactic white dwarf confusion foreground \cite{2021PhRvD.104d3019K}.
The PSD is interpolated and smoothed to mitigate numerical artifacts associated with sharp nulls in the TDI transfer functions~\cite{Nam...TransferFns} (see dashed red line in Fig.~\ref{fig:psd_curves}). 
These nulls arise from complex exponential delays in the TDI response and, although physically present, can amplify spectral leakage effects in finite data segments. 
We consider three mission durations: $T = [0.25, \, 1.5,\, 4.5]$ years, representing, respectively, an early mission, mid-stage, and nominal-phase scenario.
Since we have access only to two estimates of the PSD and foreground models \cite{2021PhRvD.104d3019K} from Ref.~\cite{FoM_inprep}, we adopt the 1.5-year PSD for mission durations of 0.25 and 1.5 years, and the 4.5-year PSD for the mission duration of 4.5 years.
We neglect the \emph{null} T channel since it suppresses the GW signals below a certain frequency~\cite{Prince:2002, Hartwig:2021tdi} and it is not going to meaningfully affect the results.

%%%%%%%%%%%%%%%%%%%%%%%%%%%%%%%%%%%%%%%%%%%%%%%%%%%%%%%%%%%%
\subsection{Pipeline Implementation}\label{subsec:pipeline}
%%%%%%%%%%%%%%%%%%%%%%%%%%%%%%%%%%%%%%%%%%%%%%%%%%%%%%%%%%%%

The numerical implementation of our analysis is based on a Python pipeline available at
\href{https://github.com/lorenzsp/EMRI-FoM}{EMRI-FoM}.

The pipeline takes as input the source-frame masses $(\mone, \mtwo)$ in solar masses, the primary dimensionless spin parameter $a$, redshift $\reds$, final eccentricity $e_f$, and the time to the end of the inspiral $T$ in years (also referred to as the time to plunge). 
The user can also specify the LISA noise PSD file, the sampling interval $\Delta t$, the number of Monte Carlo realizations $N_{\rm MC}$ over phases, sky locations, and spin inclinations, as well as whether to include the parametrized beyond-vacuum GR effects discussed previously. 
The output of the pipeline is a set of files containing the SNR and Fisher matrix results for each Monte Carlo realization.
In the following, we describe in detail the main stages of the pipeline and define the aforementioned quantities.

%%%%%%%%%%%%%%%%%%%%%%%%%%%%%%%%%%%%%%%%%%%%%%%%%%%%%%%%%%%%
\subsubsection{Trajectory generation}\label{subsubsec:trajectory}
%%%%%%%%%%%%%%%%%%%%%%%%%%%%%%%%%%%%%%%%%%%%%%%%%%%%%%%%%%%%

The first step of the pipeline is to compute the EMRI trajectory used by the waveform generator, which requires detector-frame masses and the initial orbital elements semi-latus rectum and eccentricity ($p_0, \, e_0$). We use the FEW~\cite{2025PhRvD.112j4023C} package to generate EMRI trajectories and waveforms in eccentric, equatorial orbits in Kerr~\cite{2025PhRvD.112j4023C}.
The input source-frame masses are converted into detector-frame masses using the standard cosmological relation $\monedet = (1+\reds) \mone$ and $\mtwodet = (1 + \reds) \mtwo$. 
The luminosity distance is computed from the redshift using a Planck18 cosmology~\cite{2020A&A...641A...6P, 2024arXiv240315526L}.
Since we consider equatorial orbits, the inclination is fixed, and prograde and retrograde orbits are determined by the sign of the dimensionless spin $a$ (positive for prograde and negative for retrograde).

Initial orbital parameters $(p_0,e_0)$ are obtained by integrating the orbital elements backward from the separatrix $p_s(a,e_f)$ (plus a small buffer of $2\times 10^{-3}$, enforced by the model's domain of validity), defined at the final eccentricity $e_f$. The inspiral equations {of motion}, including radiation-reaction effects, are solved using the \texttt{KerrEccEqFlux} model with adaptive step-size control and an absolute tolerance of $10^{-13}$.

For systems with parametrized deviations from vacuum GR, we employ the modified module \texttt{KerrEccEqFluxPowerLaw}, which implements the model in Eq.~\eqref{eq:modifiedtorque} for trajectory evolution. In this case, we restrict to circular orbits, for which the energy and angular momentum losses satisfy the balance relation
$\dot{E} = \Omega_\phi \dot{L}$,
where $\Omega_\phi$ is the azimuthal orbital frequency~\cite{apostolatos1993gravitational, CircularOrbitsOri}. 
For the $n_r=1$ case, we also performed a comparison with the fully relativistic trajectory, implemented in the \texttt{KerrCircEqFluxScalar} class and described in Appendix~\ref{appendix:scalar_charge}.

%%%%%%%%%%%%%%%%%%%%%%%%%%%%%%%%%%%%%%%%%%%%%%%%%%%%%%%%%%%%
\subsubsection{Waveform Generation}
% %%%%%%%%%%%%%%%%%%%%%%%%%%%%%%%%%%%%%%%%%%%%%%%%%%%%%%%%%%%%

The waveform parameter space is given by
\begin{equation}\label{eq:emriparameters}
\theta =
[\monedet, \mtwodet, a, p_0, e_0, \lumdist,
\underbrace{\Omega_S, \Omega_K, \Phi_0}_{\text{Monte Carlo}},
A, n_r] ,
\end{equation}
where the parameters
$[\monedet, \mtwodet, a, p_0, e_0, d_L, A, n_r]$
are derived from the input quantities as described above. The remaining parameters are sampled randomly. 
The sky position
$\Omega_S = (\theta_S,\phi_S)$ and the Kerr spin orientation
$\Omega_K = (\theta_K,\phi_K)$ in the Solar System Barycenter are drawn uniformly over the sphere, while the initial azimuthal and radial orbital phases
$\Phi_0 = (\Phi_{\phi,0},\Phi_{r,0})$ are sampled uniformly in $[0,2\pi]$.

The FEW package provides the plus and cross polarizations $h_+(t)$ and $h_\times(t)$ in the Solar System Barycenter frame. These are projected onto the LISA TDI 2.0 A and E channels using \texttt{fastlisaresponse}~\cite{2022PhRvD.106j3001K}, which accounts for the LISA orbital motion, unequal arm lengths, and the detector. Our choice of LISA orbits introduces small correlations between the A and E channels, which are neglected in this analysis. 
As shown in Ref.~\cite{2022PhRvD.106j3001K}, this approximation leads to only marginal changes in the SNR and parameter uncertainties, even for observation times of up to 4 years.

We include waveform harmonics down to a relative mode-amplitude threshold of $10^{-5}$, ensuring that the residual signal is of the order of $10^{-5}$ (see \cite{2025PhRvD.112j4023C} for further details on the mode selection).

% The waveform parameter space is given by

The pipeline is executed on Graphics Processing Units (GPUs), and the results presented here were obtained using an NVIDIA A100 with 41~GB of memory. This memory capacity is required to store the long-duration waveforms and intermediate quantities for the Fisher matrix calculation.
To optimize memory usage, we adopt different durations and sampling intervals:
\begin{align}
\Delta t &= 0.6~\mathrm{s}
&& \text{for } \mone = 5\times10^4,\,10^5\,M_\odot, \\
\Delta t &= 10~\mathrm{s}
&& \text{for } \mone = 10^6\,M_\odot, \\
\Delta t &= 100~\mathrm{s}
&& \text{for } \mone = 10^7\,M_\odot .
\end{align}
For beyond-vacuum GR systems observed over 4.5 years, we use $\Delta t=10$~s for $m_1=10^6\,M_\odot$ and $\Delta t=3$~s for $m_1=10^5\,M_\odot$. 
This is the maximum memory capacity allowed and we checked that our results are insensitive to these choices.
%%%%%%%%%%%%%%%%%%%%%%%%%%%%%%%%%%%%%%%%%%%%%%%%
\subsubsection{Calculating Detection and Inference proxies}
%%%%%%%%%%%%%%%%%%%%%%%%%%%%%%%%%%%%%%%%%%%%%%%%
A complete end-to-end LISA data-analysis pipeline for EMRIs/IMRIs is not yet available.
So we adopt a simplified approach to capture the essential aspects of data analysis, 
which is sufficient for the purpose of this work, i.e., to define illustrative instrumental performance criteria. 
We consider one EMRI or IMRI present in the data and assume that all the other detectable LISA source types have been identified and subtracted from the data. 

We adopt two standard diagnostics that capture the essential aspects of data analysis: 
(i) \textit{Detection}, quantified through the optimal matched-filter SNR, and 
(ii) \textit{Inference}, quantified through Fisher-matrix parameter uncertainties. 
These quantities directly inform higher-level analyses such as hypothesis testing and population studies.

The optimal matched-filter SNR is computed using the noise-weighted inner product,
\begin{equation}
\rho_C^2 (\theta) = 4 \int_{f_{\rm min}}^{f_{\rm max}} 
\frac{|\tilde{h}_C(f; \theta)|^2}{S_{n}(f)} \, df ,
\label{eq:snr_new}
\end{equation}
where $\tilde{h}_C(f;\theta)$ is the Fourier transform of the waveform in TDI channel $C$ for source parameters $\theta$. 
The one-sided noise power spectral density (PSD) $S_{n}(f)$ is the same for the two TDI channels $C=A,E$ and is a combination of instrumental noise and the galactic white dwarf confusion foreground as described in Sec.~\ref{subsec:instrument_constraint}.

The integration bounds are determined from the harmonic content of the waveform, enlarged by a 1\% safety margin. 
Equation~\eqref{eq:snr_new} assumes a stationary, Gaussian noise process; extensions to more realistic noise models are deferred to future work~\cite{burke_mind_the_gap}.
In practice, Eq.~\eqref{eq:snr_new} is evaluated using discrete Fourier transforms of tapered time-domain waveforms. 
We apply a one-day Tukey window prior to transformation to suppress spectral leakage. 

For each Monte Carlo realization over sky location $\Omega_S$, spin orientation $\Omega_K$, and initial phases $\Phi_0$, we compute
\begin{equation}
\rho = \sqrt{\sum_{C=A,E} \rho_C^2},
\end{equation}
and report the median over $N_{\rm MC}$ realizations,
\begin{equation}
\snr = 
\underset{\Omega_S,\Omega_K,\Phi_0}{\mathrm{median}} \left[ \rho \right],
\end{equation}
which reduces sensitivity to orientation-driven outliers.
To estimate parameter uncertainties, we use the Fisher information matrix~\cite{Finn:1992wt,2008PhRvD..77d2001V}, valid in the large-SNR limit under the same Gaussian, stationary noise assumptions:
\begin{align}
\Gamma^C_{ij} 
&= \left\langle \frac{\partial h_C}{\partial \theta_i} \middle| 
\frac{\partial h_C}{\partial \theta_j} \right\rangle \\
&= 4 \Re \int_{f_{\rm min}}^{f_{\rm max}} 
\frac{\partial_{\theta_i} \tilde{h}_C^*(f)\,
      \partial_{\theta_j} \tilde{h}_C(f)}
     {S_n(f)} \, df ,
\label{eq:fisher_new}
\end{align}
with corresponding covariance matrix $\Sigma = \qty(\sum_C \Gamma^C)^{-1}$ and 1-$\sigma$ uncertainties
\begin{equation}
\sigma_{\theta_i} = \underset{\Omega_S,\Omega_K,\Phi_0}{\mathrm{median}} \left[\sqrt{\Sigma_{ii}}\right].
\end{equation}
Waveform derivatives are computed numerically using the \texttt{StableEMRIFisher} package~\cite{Kejriwal_StableEMRIFisher_2026}, which implements adaptive central finite differences. 
We employ 4th-order schemes for vacuum inspirals and 8th-order schemes when including parametrized GR deviations, which require higher numerical precision.

We reparameterize the eccentricity via $e \rightarrow \log e$, which allows us to numerically differentiate in the low-eccentricity regime while ensuring positivity.
After inversion in logarithmic space, uncertainties are mapped back to the physical parameter basis using the appropriate Jacobian transformation. 
A Jacobian transformation is also used to convert detector-frame into source-frame mass measurements.

As for the SNR, we compute the Fisher matrix for each Monte Carlo realization and report the median fractional uncertainties:
\[
\sigma_{\mone}/\mone, \,
\sigma_{\mtwo}/\mtwo, \,
\sigma_{a}/a, \,
\sigma_{e_0}/e_0, \,
\sigma_{d_L}/d_L, \,
\sigma_A
\]
as well as the sky-localization area,
\begin{equation}
\Delta\Omega_S = 
2\pi \sin\theta_S 
\sqrt{\det \Sigma_S}
\left( \frac{180}{\pi} \right)^2 
\; [\mathrm{deg}^2],
\end{equation}
computed from the $2\times2$ submatrix $\Sigma_S$ associated with $\Omega_S = (\theta_S,\phi_S)$~\cite{cutler_ang_res}.

The computational cost of the pipeline is displayed in Table~\ref{tab:timings} for an NVIDIA A100 on the \href{https://doc.spider.surfsara.nl/en/latest/Pages/gpu_on_spider.html}{Spider Cluster} for the results presented in Sections~\ref{subsec:baseline_science}-\ref{subsec:mission_duration}.
We note that the timing includes diagnostic plots and checks and the 1000 Monte Carlo realizations.
The timing results exhibit a pronounced separation between the detection and inference stages. 
Detection analyses over 580 systems take a total of 1.9 days, with a median of 4 minutes per system, ranging from a minimum of 2 to a maximum of 10.4 minutes.
The inference analyses over 216 systems are substantially more expensive, with a total of 8.9 days and a median of 33 minutes per system, from a minimum of 4.6 minutes up to a maximum of 377 minutes. 
Consequently, population-scale studies will be dominated by inference expense which can be mitigated by running on different GPUs in parallel. 

\begin{table}
\centering
\begin{tabular}{lcc}
\hline
Metric & Detection & Inference \\
\hline
Number of Systems & 580 & 216 \\
Total runtime [days] & 1.93 & 8.87 \\
Median runtime [minutes] & 4 & 33 \\
Minimum runtime[minutes] & 2 & 4.6 \\
Maximum runtime [minutes] & 10.4 & 377 \\
\hline
\end{tabular}
\caption{Table of computational cost of the pipeline for the detection and inference analyses. 
Each system is defined as a unique combination of primary mass, secondary mass, primary spin, final eccentricity, time to plunge, and redshift $(\mone, \mtwo, a, e_f, T, \reds)$ for which $10^3$ Monte Carlo realizations of sky position, spin orientation, and initial phases are performed to obtain the median SNR (Detection) and parameter uncertainties (Inference).
The runtime is measured on an NVIDIA A100 GPU with 41 GB of memory.}
\label{tab:timings}
\end{table}

%%%%%%%%%%%%%%%%%%%%%%%%%%%%%%%%%%%%%%%%%%%%%
\section{Results}\label{sec:results}
%%%%%%%%%%%%%%%%%%%%%%%%%%%%%%%%%%%%%%%%%%%%%

The results of our study are organized around a progress from conservative to more ambitious science reach.
We first establish what LISA can achieve with the nominal instrument and a conservative three-month observation window (Sections~\ref{subsec:degradation_framework}--\ref{subsec:baseline_science}), then quantify how those capabilities improve with the full mission duration (Section~\ref{subsec:mission_duration}), and finally assess probes of beyond-vacuum GR with the complete 4.5-year baseline (Section~\ref{sec:InfBGR}).
For each science capability, we provide an example performance criterion quantifying the maximum allowable degradation before that capability is compromised.
%%%%%%%%%%%%%%%%%%%%%%%%%%%%%%%%%%%%%%%%%%%%%
\subsection{Instrumental Degradation Framework}\label{subsec:degradation_framework}
%%%%%%%%%%%%%%%%%%%%%%%%%%%%%%%%%%%%%%%%%%%%%

A \emph{Requirement} is a clear, testable statement that a system shall satisfy in order to meet an identified stakeholder need.
Requirements must be expressed using mandatory language (``shall''), be measurable with quantitative criteria, and remain traceable to the underlying science objectives~\cite{jacksonRequirementsAnatomy2019,visureRequirementsGuide,nasaSEH2016,koelschRequirements2016,ecssEHB40A,ecssEST40C,ecssEST1006C,ecssD0001C}.

The science objectives associated with EMRIs and IMRIs can only be achieved if these sources are both detected and accurately characterized.
However, requirements derived directly from the expected astrophysical population would be strongly affected by large uncertainties in source rates and distributions.
For this reason, we formulate requirements in terms of instrumental performance and its impact on detection and parameter estimation, rather than on absolute source counts.

We adopt the reference noise power spectral density $S_n(f)$ as the baseline instrumental configuration that defines reference values for the $\snr^{\rm ref}$ and parameter uncertainties $\sigma^{\rm ref}$.
Instrumental degradation is parameterized by a multiplicative factor $d$ applied to the noise PSD,
\begin{equation}
S_n(f) \rightarrow d\,S_n(f).
\end{equation}
Under this transformation, the detection and inference performance scale as 
\begin{equation}
\snr(d) = \frac{\snr^{\rm ref}}{\sqrt{d}},
\qquad
\sigma(d) = \sigma^{\rm ref}\sqrt{d}.
\label{eq:degradation_scaling}
\end{equation}
These scalings follow directly from the noise-weighted inner products entering the SNR and Fisher matrix definitions (Eqs.~\ref{eq:snr_new}--\ref{eq:fisher_new}). 
In reality, the impact of degradation may be more complex, depending on specific noise features and their interplay with signal characteristics.
In particular, the confusion noise estimated at 1.5 and 4.5 years with the reference PSD might not be affected by a degradation in sensitivity.
For the purpose of our work we ignore this effect and we apply the degradation factor to the overall PSD envelope assuming this leads to more conservative results.
This allows us to use the scalings of Eq.~\eqref{eq:degradation_scaling} to provide a tractable, instrument-facing framework.

Based on this, we provide examples of criteria that define acceptable performance. 
Requiring that the SNR does not decrease by more than 30\% gives approximately
\begin{equation}
\snr(d) \geq 0.7\,\snr^{\rm ref}
\quad \Rightarrow \quad
d \leq 2 \, .
\end{equation}
The SNR sets the detection threshold $\snr_{\rm det}$ relevant for population studies, since the number of detected sources scales approximately as 
$N_{\rm det} \propto (\mathrm{SNR}/\snr_{\rm det})^3$~\cite{2004CQGra..21S1595G}.
Instead, requiring that parameter uncertainties do not increase by more than 40\% gives
\begin{equation}
\sigma(d) \leq 1.4\,\sigma^{\rm ref}
\quad \Rightarrow \quad
d \leq 2 \, .
\end{equation}
Both thresholds correspond to the same degradation factor $d = 2$, which also coincides with the loss-of-one-arm scenario (see Section~4.5 of Ref.~\cite{2024arXiv240207571C}).
These statements are directly verifiable: a given instrumental configuration either satisfies the bound on $d$ or it does not.
In the following sections, we provide examples of how to formulate performance requirements and identify which sources are the most sensitive to degradation.
The reader can interactively check the results and the impact of degradation
through the \href{https://huggingface.co/spaces/lorenzsp/emri-imri-fom}{interactive website} and decide which level of degradation is best suited for their scientific goal.
We stress that these requirements are not supposed to be imposed at any level, but are examples of how to formulate a quantitative instrumental performance requirement within our framework.

%%%%%%%%%%%%%%%%%%%%%%%%%%%%%%%%%%%%%%%%%%%%%
\subsection{Baseline Science Reach}\label{subsec:baseline_science}
%%%%%%%%%%%%%%%%%%%%%%%%%%%%%%%%%%%%%%%%%%%%%

We first characterize LISA's science reach under the nominal instrument and a three-month observation window ($T = 0.25$~yr).
This baseline scenario represents the most conservative assessment: a short snapshot of the sky using only a fraction of the total mission duration.
We evaluate detection performance across cosmic history (Section~\ref{subsubsec:detection_horizon}), then use the resulting detection redshifts to evaluate parameter estimation at a fixed SNR threshold across four science probes: intrinsic masses and spin (Section~\ref{subsubsec:intrinsic_params}), orbital eccentricity (Section~\ref{subsubsec:eccentricity}), and source localization (Section~\ref{subsubsec:localization}).
Example performance criteria for each capability are provided at the end of the relevant subsection.

%%%%%%%%%%%%%%%%%%%%%%%%%%%%%%%%%%%%%%%%%%%%%
\subsubsection{Detection Horizon}\label{subsubsec:detection_horizon}
%%%%%%%%%%%%%%%%%%%%%%%%%%%%%%%%%%%%%%%%%%%%%

\begin{figure}[h!]
    \centering
    \includegraphics[width=0.95\columnwidth]{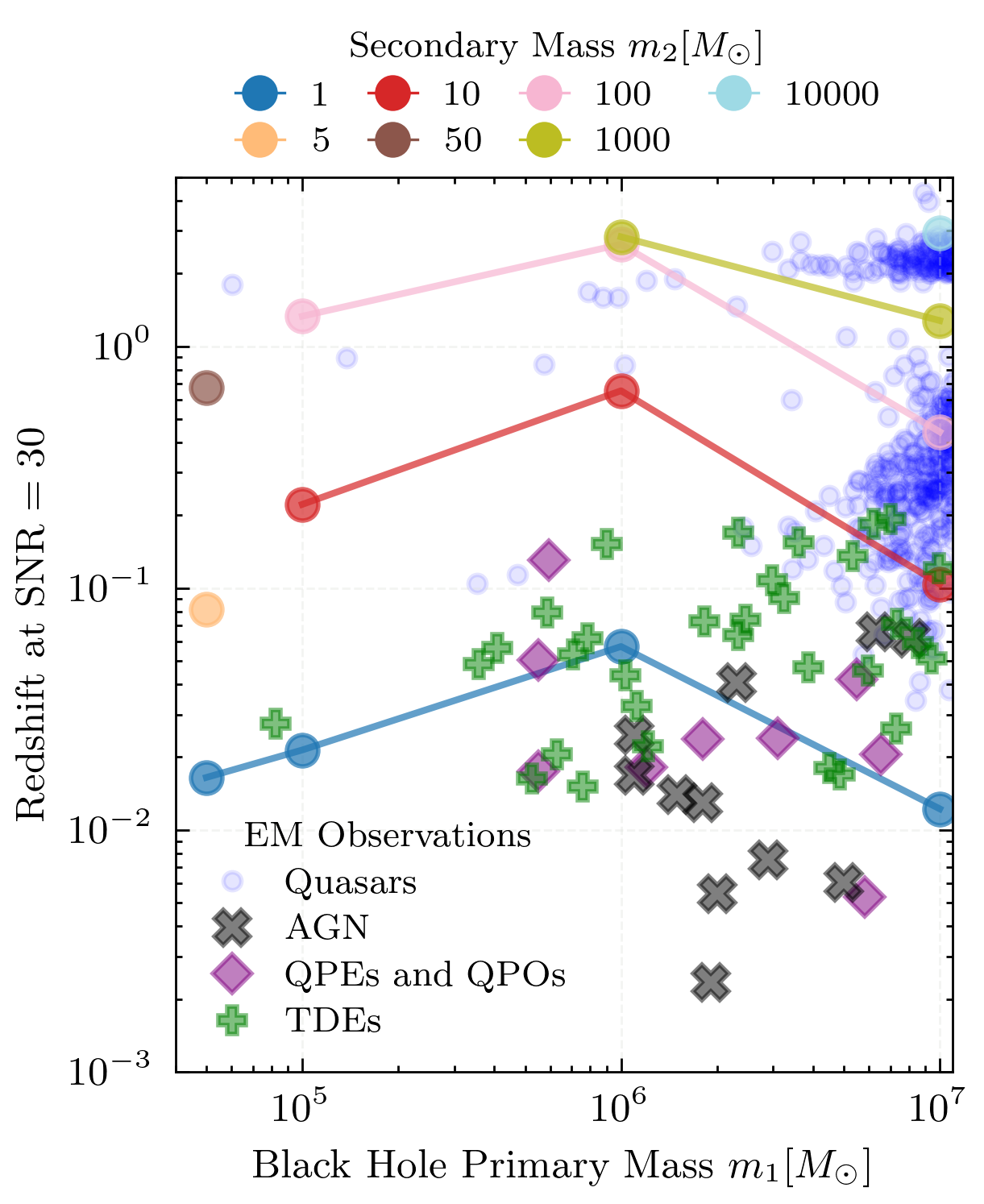}
    \caption{
    EMRI/IMRI Redshift horizon for different source-frame primary masses $\mone$ and dimensionless spin $a=+0.99$ with source-frame secondary mass $\mtwo$ for a detection threshold of $\snr = 30$.
    The observational window and the time to plunge are set to $T = 0.25$~years.
    We present estimates of Quasars (blue dots), Active Galactic Nuclei (AGN) (black cross), Quasi Periodic Eruptions and Oscillations (QPE and QPO) (violet diamond), and Tidal Disruption Events (TDEs) (green plus).
    }
    \label{fig:z_at_snr}
\end{figure}

To investigate detection performance and illustrate how a performance criterion can be defined, we compute the SNR of sources across cosmic history.
We run the pipeline with input parameters $(\mone, \mtwo, a, e_f=0, T=0.25\,\mathrm{yr}, \reds)$ across the redshift range $[10^{-3}, 10]$, sampled logarithmically in ten bins.
The corresponding luminosity distance range is approximately $4$ Mpc to $10^2$~Gpc.
The minimum redshift of $10^{-3}$ is motivated by the range of dynamical black hole mass measurements~\cite{2020ARA&A..58..257G,2025A&A...698L...9N}, while $z \sim 10$ corresponds to the epoch when the first massive black hole seeds are expected to form in assembling galactic nuclei~\cite{2010A&ARv..18..279V,2020ARA&A..58...27I,2025ApJ...983L...4F}.

We consider the 15 mass pairs $(\mone, \mtwo)$ of Fig.~\ref{fig:emri_imri_population}, each with spins $a = \pm0.99$, yielding 30 system configurations. 
We employ the noise PSD corresponding to a mission duration of $T = 1.5$~yr.
By interpolating the median SNR as a function of redshift for each $(\mone, \mtwo, a)$ configuration, we obtain the detection horizon ${z}(\snr)$ shown only for prograde orbits in Fig.~\ref{fig:z_at_snr} for $a = +0.99$ and $\snr = 30$~\cite{2025arXiv251020891S}.
Retrograde orbits are not shown but are discussed in the text. 
We encourage the reader to explore the full set of results, including retrograde orbits, through the \href{https://huggingface.co/spaces/lorenzsp/emri-imri-fom}{interactive website}.

Three features stand out. 
First, the redshift horizon increases monotonically with secondary mass $\mtwo$, since SNR scales linearly with $\mtwo$ at fixed primary mass and distance. 
Second, the horizon peaks around $\mone \sim 10^6\,M_\odot$: the primary mass sets the gravitational wave frequency, and the LISA band reaches its minimum strain sensitivity at $\sim 7.9 \times 10^{-3}$~Hz, which for $a = +0.99$ ($a = -0.99$) corresponds to a last-stable-orbit frequency from a primary of $\mone \sim 3 \times 10^6\,M_\odot$ ($\mone \sim 3 \times 10^5\,M_\odot$). 
Third, and most relevant for mission planning, the systems at the extremes of the mass-ratio range --- light secondaries ($m_2 = 1\,M_\odot$) and retrograde orbits --- have the shallowest horizons and are therefore most sensitive to sensitivity degradation.

From the horizon plot, LISA will detect EMRIs in electromagnetically observed galactic centers up to redshift $z \sim 0.01$--$0.1$. 
IMRIs, with their heavier secondaries, reach much further: LISA can assess whether high-redshift quasars host heavy IMRIs with $m_1 = 10^7\,M_\odot$, $m_2 = 10^3\,M_\odot$ out to $z \sim 1.2$.

\subsubsection*{Detection requirement}
By choosing $T = 0.25$~yr and the lightest secondary $m_2 = 1\,M_\odot$, we provide the following example requirement on detection performance:

\emph{The LISA performance shall ensure that, for every pair $\{m_1, a\}$ in the EMRI/IMRI reference population with $m_2 = 1\,M_\odot$ and $z_{\rm ref} = 0.0215$, the median SNR does not decrease by more than 30\% relative to the baseline PSD evaluation.}

\noindent Acceptance criterion: $\snr(d)/\snr^{\rm ref} \geq 0.7$ for all such sources, verified with the pipeline.
The reference redshift $z_{\rm ref} = 0.0215$ can be replaced by any redshift within the horizon, since the requirement is on the relative SNR loss.
The 30\% threshold is motivated by the cubic scaling $N_{\rm det} \propto (\snr / \snr_{\rm det})^3$: a 30\% SNR loss corresponds to roughly a factor-of-two reduction in detectable sources.
By using both spins $a = \pm0.99$ we ensure the requirement covers the full spin distribution, from the most optimistic prograde to the most pessimistic (retrograde) scenarios.
A degradation of $d = 1.5$ would reduce the SNR by $\approx 20\%$, marginally within the acceptable range; the first sources lost are EMRIs with $m_2 = 1\,M_\odot$ on retrograde orbits around $\mone = 10^7\,M_\odot$ primaries.

%%%%%%%%%%%%%%%%%%%%%%%%%%%%%%%%%%%%%%%%%%%%%
\subsubsection{Masses and spin inference}\label{subsubsec:intrinsic_params}
%%%%%%%%%%%%%%%%%%%%%%%%%%%%%%%%%%%%%%%%%%%%%

\begin{figure}
    \centering
    \includegraphics[]{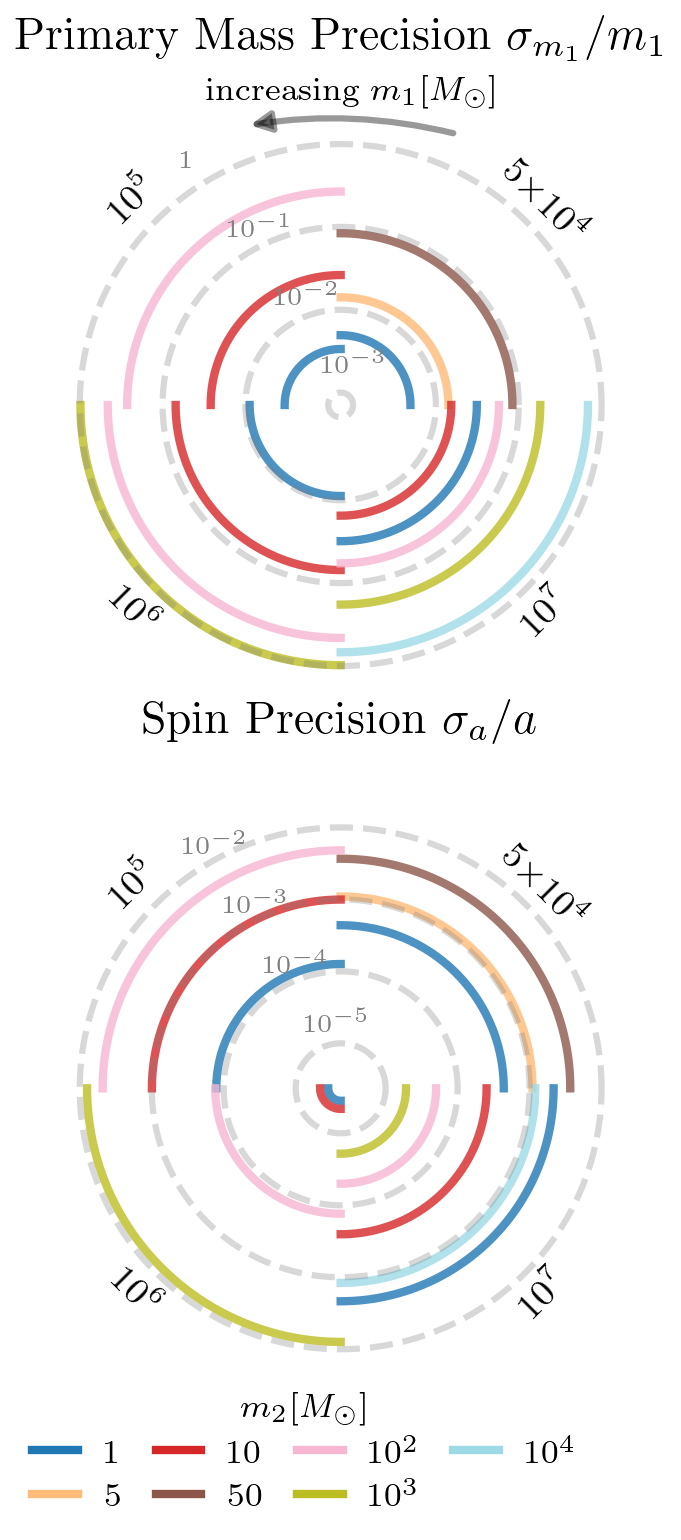}
    \caption{
    Relative measurement precision for source-frame primary mass $m_1$ (top panel) and dimensionless spin $a$ (bottom panel) as a function of component masses $\mone, \mtwo$ for prograde orbits ($a = 0.99$) at fixed median SNR of 30 and three-month observations.
    % The relative precision on the primary is dominated by the luminosity distance uncertainty; detector-frame mass precision ranges over $\sigma_{\monedet}/\monedet \approx 10^{-5}$--$10^{-3}$.
    }
    \label{fig:precision_m1_a}
\end{figure}

If the detection requirement is satisfied, we proceed to evaluate parameter estimation.
For each system configuration $(\mone, \mtwo, a)$ we place the source at the redshift $\reds(\snr = 30)$ obtained from the detection horizon analysis and run the inference pipeline with $e_f = 0$, $T = 0.25$~yr.
Systems that do not reach $\snr \geq 30$ within $\reds \in [10^{-3}, 10]$ are excluded; in practice this affects only $a = -0.99$, $m_1 = 10^7\,M_\odot$, $m_2 = 1, 10\,M_\odot$.
The full list of results is provided in Appendix~\ref{app:precision} and in Table~\ref{tab:precision_table}.
In the following, we present the results for prograde orbits.

\paragraph{Mass measurements.}
The top panel of Fig.~\ref{fig:precision_m1_a} shows the relative uncertainty on the source-frame primary mass $\sigma_{\mone}/\mone$.
Precision improves for smaller secondary masses with precision of $\sigma_{m_1}/m_1 \sim 0.34\%$ for low-mass systems ($m_1 = 10^5\,M_\odot$, $m_2 = 1\,M_\odot$), degrading to $\sim 98\%$ for high-mass, high-mass-ratio systems ($m_1 = 10^6\,M_\odot$, $m_2 = 10^3\,M_\odot$).
The secondary mass follows the same trend, from $\sim 0.3\%$ at the lowest mass ratios to $\sim 77\%$ for $q = 10^{-3}$.
These source-frame uncertainties are dominated by the luminosity distance uncertainty, which enters through the Jacobian transformation as $\sigma_{d_L} / (1+z)\,\mathrm{d}z/\mathrm{d}d_L$.
Detector-frame masses are measured with substantially higher precision, in the range $\sim 10^{-5}$--$10^{-1}$, representing an improvement of one to two orders of magnitude relative to source-frame quantities.
Precision degrades slightly at the edges of the LISA band (lowest and highest $\mone$), while retrograde orbits worsen $\sigma_{\monedet}/\monedet$ by approximately an order of magnitude compared to prograde.

\paragraph{Spin measurements.}
The bottom panel of Fig.~\ref{fig:precision_m1_a} shows the dimensionless spin precision $\sigma_a/a$.
Owing to the strong dependence of the EMRI waveform phase on the Kerr spin~\cite{PhysRevD.102.124054}, spin is measured with accuracy, typically in the range $10^{-6}$--$10^{-2}$.
All systems achieve sub-percent spin precision, with the tightest constraints obtained for low-mass secondaries ($m_2 \leq 10\,M_\odot$) largely independent of primary mass.
This accuracy enables precision tests of the Thorne limit ($a \lesssim 0.998$) and constraints on black hole formation and growth scenarios.
The largest uncertainties occur for IMRIs with $m_2/m_1 = 10^{-3}$; the tightest measurement, $\sigma_a/a \approx 4\times  10^{-6}$, is achieved for $m_1 = 10^6\,M_\odot$ and $m_2 = 1\,M_\odot$.
Retrograde orbits are worse by up to three orders of magnitude: for $m_1 = 10^7\,M_\odot$ the spin becomes effectively unconstrained, while the best retrograde constraint reaches $\sigma_a/a \approx 3 \times 10^{-3}$ for $m_1 = 5 \times 10^4$--$10^5\,M_\odot$ and light secondaries.

\paragraph{Caveat: secondary spin.}
In this analysis, we neglect the contribution of the secondary spin to the orbital phase.
This effect enters the radiation-reaction fluxes at second-order in the mass ratio and contributes only $\mathcal{O}(1)$ radians over the full inspiral, compared to the dominant $\mathcal{O}(m_1/m_2)$ phase evolution~\cite{2024PhRvD.109l4048B}.
For EMRIs, previous studies on circular orbits around non-rotating primaries show that the secondary spin is not directly detectable and has negligible impact on other parameters; we therefore expect our EMRI results to be unaffected.
For IMRIs, however, including the secondary spin leads to a degradation of intrinsic parameter measurements by approximately one order of magnitude due to strong parameter correlations~\cite{2024PhRvD.109l4048B}, so the IMRI mass and spin measurements reported here should be regarded as optimistic.
A reassessment will be required once waveform models including secondary spin become available.

%%%%%%%%%%%%%%%%%%%%%%%%%%%%%%%%%%%%%%%%%%%%%
\subsubsection{Orbital Eccentricity}\label{subsubsec:eccentricity}
%%%%%%%%%%%%%%%%%%%%%%%%%%%%%%%%%%%%%%%%%%%%%

\begin{figure}[h!]
    \centering
    \includegraphics[width=0.99\columnwidth]{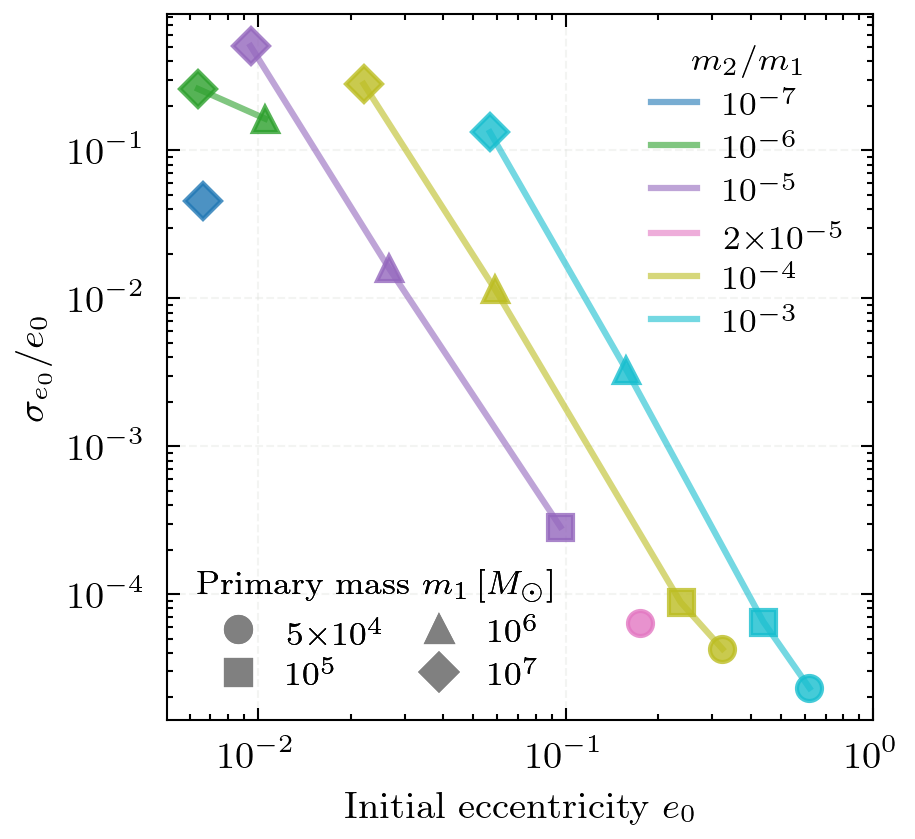}
    \caption{Relative precision on the initial orbital eccentricity $\sigma_{e_0}/e_0$ as a function of $e_0$ for different primary masses $\mone$ (indicated by the markers) and mass ratios (indicated by line colors).}
    \label{fig:precision_eccentricity}
\end{figure}

The orbital eccentricity of EMRI/IMRIs also encodes key information about their formation channels. Systems formed through two-body relaxation in dense stellar cusps are expected to retain significant eccentricity at plunge~\cite{2025arXiv250902394M}, while those formed via tidal capture or in accretion disks are expected to be nearly circular~\cite{2025arXiv250900469S}.

Across all sources in Fig.~\ref{fig:emri_imri_population}, we set $e_f = 10^{-2}$ and $T = 0.25$~yr to study their measurement precision for the initial eccentricity $e_0$.
Fixing $e_f$ ensures that all systems with the same spin follow identical orbital trajectories in the semi-latus rectum--eccentricity ($p$--$e$) plane; with this choice IMRIs have larger initial eccentricities than EMRIs.

Figure~\ref{fig:precision_eccentricity} shows $\sigma_{e_0}/e_0$ as a function of $e_0$ for different combinations of $\mone$ and $\mtwo/\mone$.
The precision depends strongly on mass ratio, and for fixed mass ratio scales approximately as $\sigma_{e_0}/e_0 \propto e_0^2$ for mass ratios $10^{-3}$, $10^{-4}$, and $10^{-5}$.
This relation can be explained by the direct proportionality between relative measurement precision and the square root of the mismatch, which was shown in Fig.~18 of \cite{2025PhRvD.112j4023C} to scale as $(e_0)^2$.
This is expected because higher eccentricities produce richer harmonic content that breaks parameter degeneracies.
The relative precision $\sigma_{e_0}/e_0$ ranges from 50\% for $m_1=10^7\,M_\odot,\,m_2=100\,M_\odot$ to $0.002\%$ for $m_1=5\times 10^4\,M_\odot,\,m_2=50\,M_\odot$.
The achieved precision should be sufficient to discriminate between wet and dry formation channels (see Table~I of Ref.~\cite{2025arXiv250900469S}), providing a powerful probe of black hole growth and stellar dynamics in galactic nuclei.

%%%%%%%%%%%%%%%%%%%%%%%%%%%%%%%%%%%%%%%%%%%%%
\subsubsection{Source Localization and Distance}\label{subsubsec:localization}
%%%%%%%%%%%%%%%%%%%%%%%%%%%%%%%%%%%%%%%%%%%%%

\begin{figure}
    \centering
    \includegraphics[]{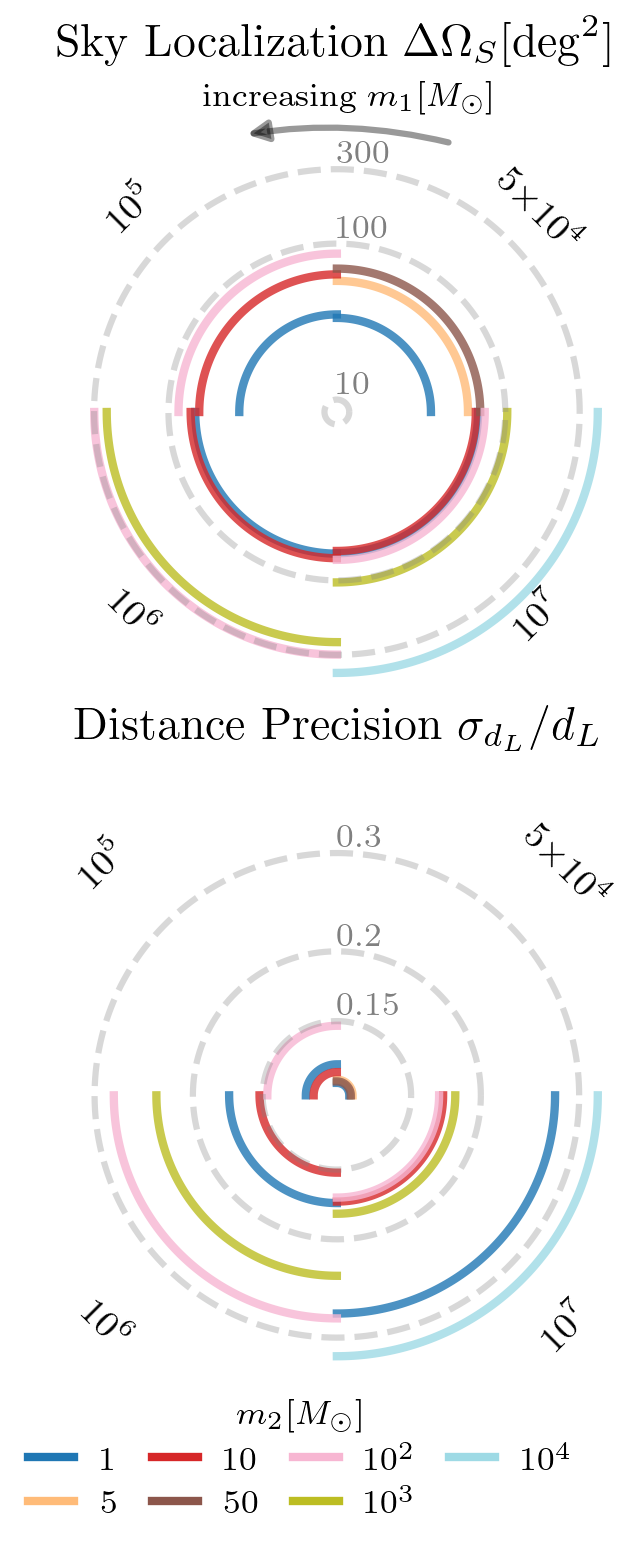}
    \caption{Sky localization uncertainty $\Delta\Omega_S$ in square degrees (top panel) and relative luminosity distance precision $\sigma_{d_L}/d_L$ (bottom panel) for prograde EMRI/IMRI systems ($a = 0.99$) at fixed median SNR of 30 and three-month observations.
    In the top panel, the sky localization for primary mass $\mone = 10^7\,M_\odot$ and $\mtwo = 1, 10, 10^2 \,M_\odot$ are similar, so the curves overlap.
    In the bottom panel, the distance precision for primary masses $\mone = 5\times 10^4\,M_\odot$ are similar, so the three curves overlap.
    }
    \label{fig:precision_extrinsic}
\end{figure}

Accurate source localization, determined by the sky position $\Omega_S$ and luminosity distance $d_L$, is essential for multi-messenger astronomy: it enables targeted electromagnetic follow-up of EMRI/IMRI host galaxies and allows these systems to be used as standard sirens for cosmology~\cite{2021MNRAS.508.4512L}.

Figure~\ref{fig:precision_extrinsic} shows the localization performance at fixed median SNR of 30 and three-month observations.
LISA can typically localize EMRIs to within $30$--$500\,\mathrm{deg}^2$, with the best performance for systems with $\mone \sim 10^5\,M_\odot$ and low-mass secondaries.
% These systems remain in band for longer durations, accumulating phase information that breaks sky-position degeneracies; the dominant driver is LISA's orbital motion, which induces Doppler modulation and annual parallax effects that encode the source direction.
Luminosity distance is measured with $\sigma_{d_L}/d_L \sim 11$--$32\%$ across the parameter space, with weak dependence on source configuration.
In this work, we neglect the impact of peculiar velocities and gravitational lensing, which contribute at the few-percent level to the redshift uncertainty at low redshift ($z = 0.1$) and at the $\sim10\%$ level for $z \gtrsim 1$ \cite{2021PhRvD.103h3526S}.
Compared to previous works \cite{2017PhRvD..95j3012B} where $10 \mathrm{deg}^2 \gtrsim \Delta \Omega_S $, the sky localizations are substantially larger.
This is due to the shorter duration (3 months), the systems considered (EMRI to IMRIs) and the considered orbits (equatorial circular).
These large uncertainties on the sky localization would make multi-messenger science particularly challenging. 
However, these results are for 3 months observations and equatorial circular orbits.
As we will see in Sec.~\ref{subsec:mission_duration}, a longer observational baseline would allow sky localization below 10 square degrees.
% This level of precision, together with a redshift measurement obtained from an electromagnetic counterpart, could enable constraints on cosmological parameters \cite{2021MNRAS.508.4512L}. 

\subsubsection*{Inference requirement}
The three subsections above (Sections~\ref{subsubsec:intrinsic_params}--\ref{subsubsec:localization}) together allow us to define the following example of the inference requirement:

\emph{The LISA performance shall ensure that, for every source in the EMRI/IMRI reference population, the relative uncertainties on source-frame component masses $\sigma_{m_{1,2}}/m_{1,2}$, primary spin $\sigma_a/a$, luminosity distance $\sigma_{d_L}/d_L$, and sky localization $\Delta\Omega_S$ do not increase by more than 40\% relative to the baseline evaluation.}

\noindent Acceptance: $\sigma(d)/\sigma_{\rm ref} \leq 1.4$ for each parameter, verified with the pipeline.
Under the degradation scaling of Eq.~\eqref{eq:degradation_scaling}, this corresponds to $d \leq 2$.
The sources most sensitive to degradation are IMRIs, whose shorter in-band duration concentrates most of the signal power near plunge, making their parameter estimation more susceptible to noise increases than the extended EMRI signal.

%%%%%%%%%%%%%%%%%%%%%%%%%%%%%%%%%%%%%%%%%%%%%
\subsection{Impact of Mission Duration}\label{subsec:mission_duration}
%%%%%%%%%%%%%%%%%%%%%%%%%%%%%%%%%%%%%%%%%%%%%

\begin{figure}[hbtp!]
    \centering
    \includegraphics[width=0.95\columnwidth]{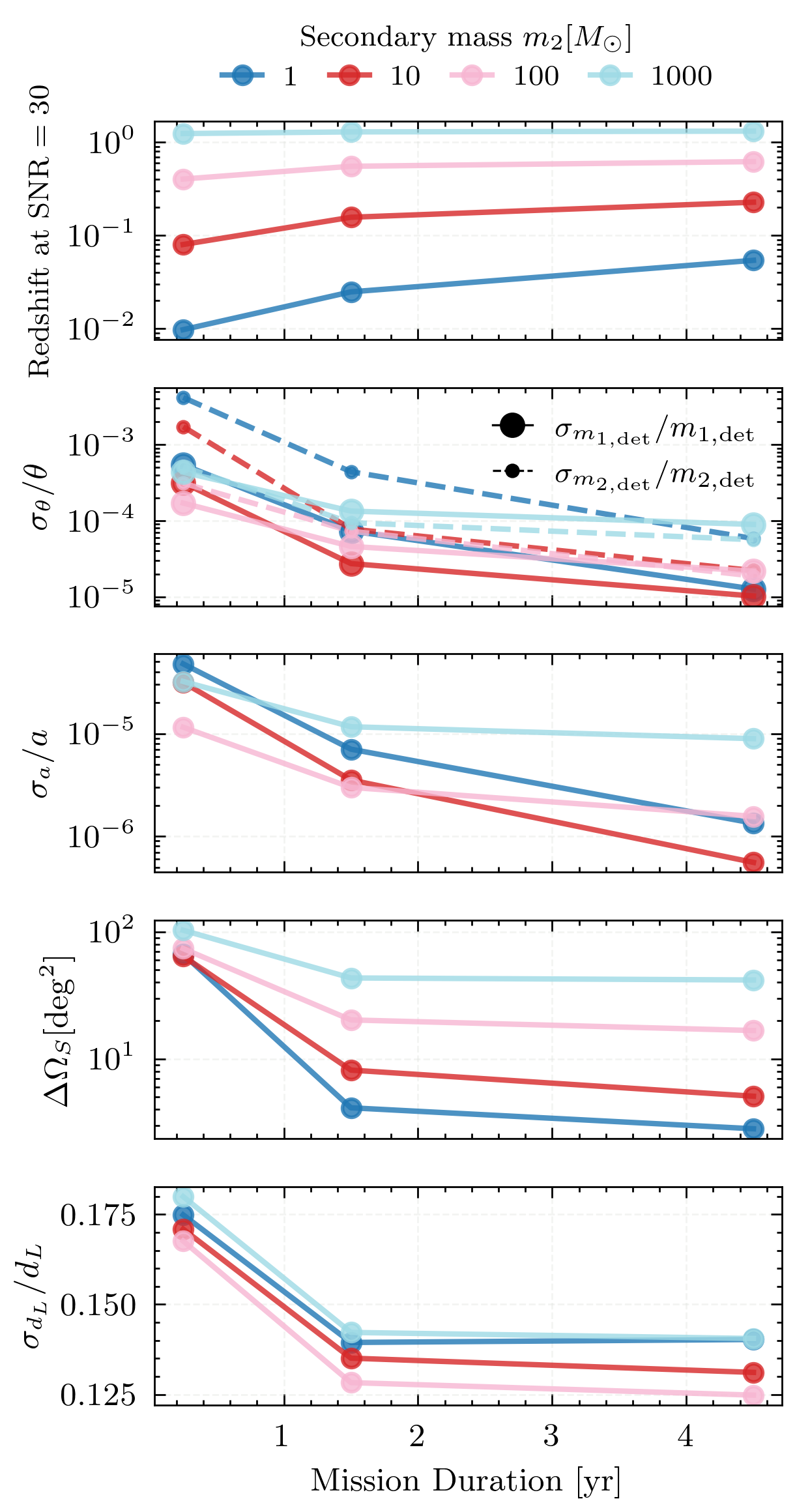}
    \caption{
    Redshift horizon and parameter measurement precision as a function of mission duration equal to time to plunge for systems with primary mass $\mone = 10^7\,M_\odot$ and dimensionless spin $a = +0.99$ for fixed median $\snr = 30$.
    \textit{From top to bottom}: redshift horizon; detector-frame mass uncertainties $\sigma_{\monedet}/\monedet$ and $\sigma_{\mtwodet}/\mtwodet$; sky localization uncertainty $\Delta\Omega_S$; and relative luminosity distance uncertainty $\sigma_{d_L}/d_L$.
    }
    \label{fig:precision_tpl}
\end{figure}

The three-month baseline of the previous section is deliberately conservative.
In practice, the mission duration changes what EMRI/IMRI science is possible, through three effects.
First, longer observations accumulate more orbital cycles, directly increasing the SNR.
Second, additional phase information from a longer signal breaks parameter degeneracies, improving mass and spin precision.
Third, an extended observation baseline amplifies the Doppler modulation and parallax effects that encode sky position, substantially improving localization.
These effects are more impactful for EMRIs than IMRIs, since they are longer-lived and accumulate SNR uniformly in time. 
Nonetheless, IMRIs' detection horizon still grows with mission duration.

We quantify these effects by varying the time to plunge together with the mission duration over $0.25$, $1.5$, and $4.5$~yr and computing detection horizons and parameter estimation precision for systems with mass ratios $m_2/m_1 \leq 10^{-3}$ at fixed SNR.
Fixing the SNR allows us to isolate the improvement in measurement precision arising from waveform information only rather than from changes in signal strength.
Figure~\ref{fig:precision_tpl} shows the results for a representative system with $\mone = 10^7\,M_\odot$, $a = +0.99$, and zero eccentricity.

For the $m_2 = 1\,M_\odot$ system, extending observation from 0.25 to 4.5~yr increases the detection horizon redshift by a factor of $\sim 4$.
For parameter estimation at fixed $\snr = 30$, the gains from the full 4.5-year mission are substantial:
\begin{itemize}
    \item \textit{Redshift horizon} increases by factors of $\sim 2$--$4$ relative to the 0.25-yr baseline, with the largest gains for low-mass-ratio systems that remain in-band longer;

    \item \textit{Mass precision} improves by factors of $\sim 2$--$3$, with detector-frame measurements reaching $\sigma_{m_1,\det}/m_{1,\det} \sim 10^{-5}$ for optimal configurations; the improvement exceeds one order of magnitude for secondary masses $m_2 = 1, 10, 100\,M_\odot$, and amounts to a factor of a few for $m_2 = 10^3\,M_\odot$;

    \item \textit{Sky localization} improves by factors of $\sim 10$--$100$, achieving precision of $\Delta\Omega_S < 10\,\mathrm{deg}^2$ for $m_2 = 1, 10\,M_\odot$;
    %  the improvement exceeds one order of magnitude for , and is a factor of a few for $m_2 > 10\,M_\odot$;

    \item \textit{Luminosity distance} precision improves by at most a factor of 2, occurring for $m_2 = 1\,M_\odot$;

    \item \textit{Spin} {precision} improves by up to two orders of magnitude for the most extreme mass ratios ($m_2 = 1$--$10\,M_\odot$), while IMRI systems show modest improvement.
\end{itemize}

Retrograde orbits with $a = -0.99$ show similar qualitative trends with systematically worse absolute precision.
For prograde systems with $e_f = 0.01$, longer observations also substantially improve eccentricity constraints.
The relative precision $\sigma_{e_0}/e_0$ improves by approximately one order of magnitude for $m_2 = 1$--$10\,M_\odot$ and nearly two orders of magnitude for $m_2 = 100$--$1000\,M_\odot$, enabling sharper discrimination between formation channels.

\subsubsection*{Mission duration requirement}
An example performance requirement on mission duration is:

\emph{The LISA performance shall ensure that the SNR and measurement precision at 4.5 years shall not degrade by more than 30\% and 40\%, respectively.}

\noindent Acceptance: $\sigma(d,T=4.5\mathrm{yr})/\sigma_{\rm ref}(T=4.5\mathrm{yr}) \leq 1.4$ for each parameter and $\snr(d)/\snr^{\rm ref} \geq 0.7$ for all sources, verified with the pipeline.
This requirement can then be translated into mission-duration and duty-cycle allocations.
% We recall that data gaps reduce the effective SNR approximately as a multiplicative factor of 0.82 and do not cause complete signal loss for EMRIs, owing to their quasi-monochromatic character during the early inspiral. IMRIs, which evolve more rapidly, are more sensitive to gaps in temporal coverage.

%%%%%%%%%%%%%%%%%%%%%%%%%%%%%%%%%%%%%%%%%%%%%
\subsection{Probing Beyond-Vacuum General Relativity}\label{sec:InfBGR}
%%%%%%%%%%%%%%%%%%%%%%%%%%%%%%%%%%%%%%%%%%%%%

\begin{figure}[hbtp!]
    \centering
    \includegraphics[width=0.95\columnwidth]{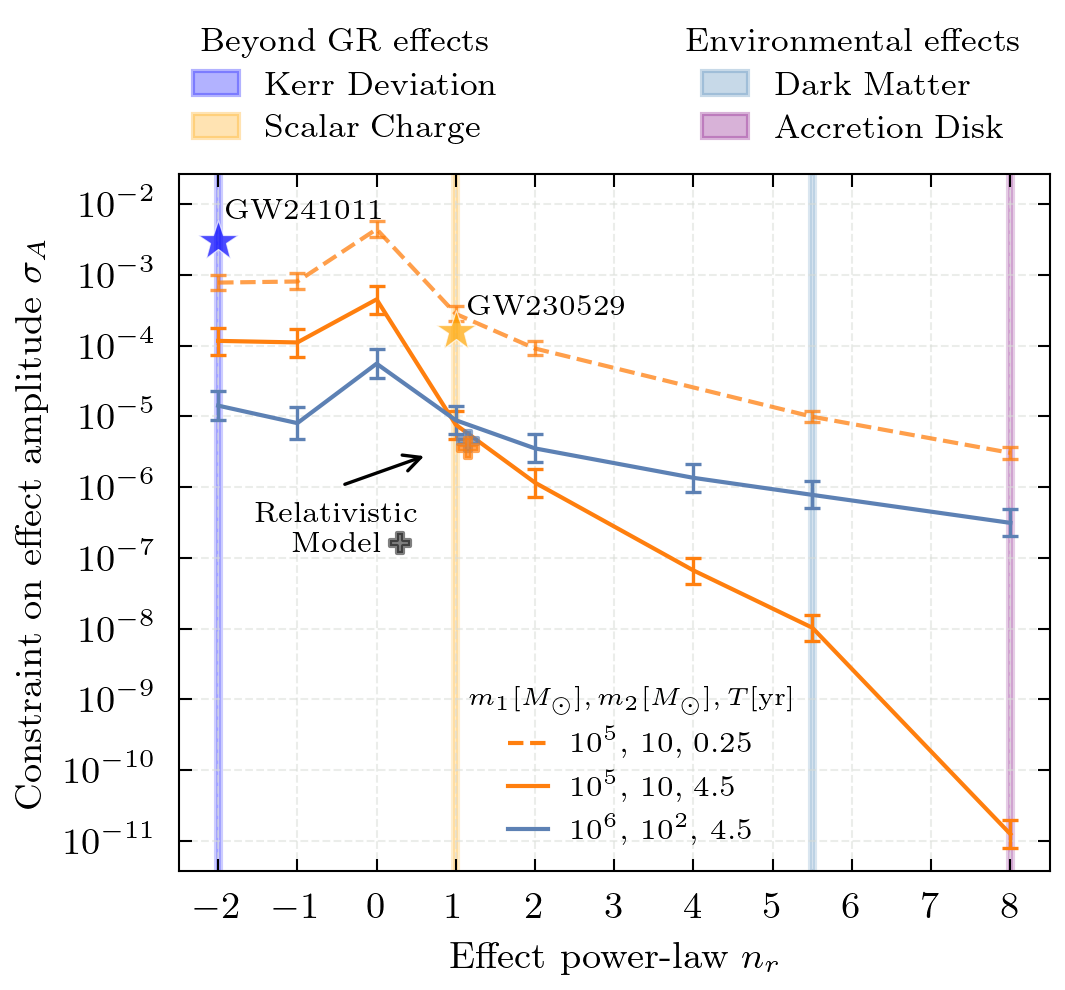}
    \caption{
    Expected constraints on deviations from vacuum General Relativity with circular extreme mass ratio inspiral observations with primary and secondary masses $(m_1, m_2)$ and observation length of $T$.
    The $y$-axis shows the measurement precision on the fractional amplitude $A$ of angular momentum loss deviating from vacuum General Relativity.
    We show two class of deviations: modifications to General Relativity, denoted by Beyond GR effects, and environmental effects.
    These deviations modify the inspiral rate with a characteristic radial dependence $\propto p^{n_r}$, shown as vertical lines.
    Any effect with an amplitude above the constraint line would be detectable by such an observation.
    Stars indicate the tightest constraints from ground-based events GW241011 and GW230529.
    For the case of scalar emission $n_r=1$, we also show the expected constraints obtained with a fully relativistic beyond GR model as plus markers.
    }
    \label{fig:precision_nr}
\end{figure}

Having established in Section~\ref{subsec:mission_duration} that the full nominal 4.5-year mission substantially improves parameter estimation, 
we now consider the extended observation time to constrain deviations from vacuum GR.
The analysis is restricted to prograde orbits with $a = +0.99$, as these configurations cause the secondary object to probe the strongest-field region around the primary.

We consider vacuum EMRIs ($A = 0$) and, for each fixed power-law index $n_r$, include $A$ as a free parameter in the inference.
Its inferred uncertainty is then interpreted as the minimum fractional deviation from vacuum that would be detectable.
We focus on two fiducial circular EMRI systems with source-frame masses $(m_1, m_2) = (10^6, \, 10^2)\,M_\odot$ and $(10^5, \, 10)\,M_\odot$, placed at redshifts $z = 0.5$ ($d_L = 2.92$ Gpc) and $z = 0.25$ ($d_L = 1.30$ Gpc), with median SNRs of 118 and 42, respectively. 
These sources correspond to two \textit{golden} EMRIs with distinct frequency evolutions, both lying within the bucket of the LISA sensitivity band. 
Since deviations from vacuum GR are expected, in most cases, to be small, these two systems can be regarded as proxies for the most favorable scenarios in which to probe such effects. 
% Consequently, any requirements derived from them should be considered conservative, as any loss of sensitivity would likely be even more detrimental to other systems. 
Future work should extend this analysis to a full population-level study~\cite{2025arXiv251017398K}. 

Figure~\ref{fig:precision_nr} summarizes the results for all physical effects ($n_r$) considered. A key trend governs the results: systems with lower primary mass place tighter constraints on $n_r > 0$, while systems with higher primary mass are more effective for $n_r < 0$. This is expected as 
lower-mass systems have a faster frequency evolution in the detector frame and therefore, for the same observation time prior to plunge, enter the observational band at larger orbital separations where the deviation amplitude $A$ at $n_r > 0$ is enhanced.
Conversely, higher-mass systems spend longer in the strong-field region and are more effective at constraining strong-field effects with $n_r < 0$.

\paragraph{Disk torques ($n_r = 8$).}
Migration torques from a surrounding gas disk, exerted by spiral density waves launched by the compact object's orbital motion, enter the angular-momentum flux with a steep radial dependence, $n_r=8$, in the fiducial $\alpha$-disk model~\cite{1973A&A....24..337S} (see Appendix~\ref{sec:AppendixDisk} for details on the mapping between the power-law amplitude and the disk parameters, e.g.\ the accretion rate). We find median measurement precision of $\sigma_A^{n_r=8}=3.3\times10^{-7}$ for the $(10^6,\,10^2)\,M_\odot$ system and $\sigma_A^{n_r=8}=1.2\times10^{-11}$ for the $(10^5,\,10)\,M_\odot$ system. For reference, $A=1.4\times10^{-6}$ for a disk with $\alpha=0.05$ and $f_{\rm Edd}=0.01$ around a $10^6\,M_\odot$ primary, with $A\propto \alpha^{-1} f_{\rm Edd}^{-3}$. Over the astrophysically relevant range $\alpha,\,f_{\rm Edd}\sim 0.01$--$0.1$, which implies that LISA will be sensitive to gaseous disk environments around lower-mass EMRIs, in agreement with previous studies~\cite{2023PhRvX..13b1035S, 2023NatAs...7..943C, 2023PhRvX..13b1035S}. 

For a shorter observation time prior to plunge, the constraints degrade by $\sim 5$ orders of magnitude for the high-mass system, as expected: the signal begins at a smaller initial radius, reducing the accumulated deviation in the angular-momentum loss relative to GR (which in this model scales as $\propto p^8$). At this level, the resulting bounds cannot exclude realistic astrophysical disk configurations, and this particular science investigation would therefore not be achievable in that observing scenario.
Finally, we note that relativistic corrections, here ignored, can enhance the torque magnitude by up to an order of magnitude in the strong-field regime~\cite{2025arXiv251002433D, 2025PhRvD.112l4068H, 2025PhRvD.112l4012H, 2026arXiv260119123D}, further improving LISA's sensitivity to these effects.

\paragraph{Dynamical friction from dark matter ($n_r = 5.5$).}

Dynamical friction~\cite{1943ApJ....97..255C, Vicente:2022ivh} from a constant-density dark matter environment imprints a radial slope $n_r = 5.5$ on the inspiral (see Appendix~\ref{sec:AppendixDM} for details).
For the same reason as for the disk torques, the lower-mass system provides tighter constraints: 
$\sigma_A^{n_r=5.5} = 1.1 \times 10^{-8}$ compared to $\sigma_A^{n_r=5.5} = 7.8 \times10^{-7}$ for the higher-mass system.

For reference, $A \approx 6.5 \times 10^{-6}$ ($6.5 \times 10^{-9}$) for $\rho_{\rm DM} = 10^{17}\,M_\odot/\mathrm{pc}^3$ ($10^{16}\,M_\odot/\mathrm{pc}^3$) and $m_1 = 10^6\,M_\odot$ ($10^5\,M_\odot$).
Dark matter overdensities formed by adiabatic contraction around massive black holes can reach densities of $\rho_{\rm DM} \lesssim 10^{19}\,M_\odot/\mathrm{pc}^3$~\cite{1999PhRvL..83.1719G, 2025arXiv251209985C} (and boson clouds grown via superradiance can be even denser~\cite{2015LNP...906.....B}), and as in disks relativistic corrections can further boost dynamical friction by orders of magnitude in the strong-field regime~\cite{2022PhRvL.129x1103C, 2025PhRvL.135u1401V, 2022PhRvD.105f1501C, 2024PhRvL.133l1404D, 2025PhRvL.134u1403D, 2023PhRvD.108h4019B}.
These results confirm that LISA can constrain the existence of dark matter structures in galactic nuclei, in agreement with previous studies~\cite{2014PhRvD..89j4059B, 2020A&A...644A.147C, Kavanagh:2020cfn, 2023NatAs...7..943C}.

\paragraph{Secondary scalar charge ($n_r = 1$).}
A broad class of scalar--tensor theories predicts that the secondary compact object acquires a scalar charge $d$, which sources dipole radiation entering at $-1$ post-Newtonian (PN) order relative to the GR quadrupole flux in the weak-field region, corresponding to $n_r = 1$ (see Appendix~\ref{appendix:scalar_charge}).
We obtain $\sigma_A^{n_r=1} = 8.8 \times 10^{-6} $ for $(m_1, m_2) = (10^6, 100)\,M_\odot$ and $\sigma_A^{n_r=1} = 7.6 \times 10^{-6} $ for $(m_1, m_2) = (10^5, 10)\,M_\odot$  from the parametrized power-law trajectory.
We also derive constraints from a fully relativistic scalar-emission model by mapping it onto the parametric formalism, which enters the weak-field regime at $-1$PN order relative to the General Relativity quadrupole flux (see Eq.~\ref{eq:A_d_relation} and Appendix~\ref{appendix:scalar_charge} for details).
This fully relativistic trajectory yields
$\sigma_A = 4.5\times 10^{-6}$
for $(m_1, m_2) = (10^6, 100)\,M_\odot$ and $T=4.5$ yr, and $\sigma_A = 3.6 \times 10^{-6} \, (1.6 \times 10^{-4})$
for $(10^5, 10)\,M_\odot$ at $T=4.5 \, (0.25)$ yr.

The relativistic flux model provides constraints which are a factor of a few smaller than the parametrized model.
This is expected because the relativistic flux model does not follow a power law with $n_r=1$, but contains a more complex dependence on radius that makes the slope effectively steeper (approximately $n_r \sim 1.15$).

For this case, a shorter observation time of $T=0.25$ yr does not worsen the constraints by so many orders of magnitude as for more negative $n_r$. 
Nonetheless, they are comparable to the ground-based constraint from the event GW230529~\cite{2024arXiv240603568S}, which set 
% $A_{\rm GW230529} \lesssim 6.4 \times 10^{-4}$ 
$A_{\rm GW230529} \lesssim 1.6 \times 10^{-4}$ 
for $n_r = 1$ (see Appendix~\ref{appendix:scalar_charge} for details).
LISA improves on this bound by roughly one order of magnitude (although note this is probing a different range of the binary parameter space). 
Also, a previous study concluded that constraints from the Einstein Telescope for a GW230529-like event should be $\sim 1$ order of magnitude better than the one set by LVK~\cite{2026PhRvD.113b3036S}, but still worse when compared with our prediction for LISA.
For a more detailed comparison see Fig.~\ref{fig:bound_delta_phi} and Appendix~\ref{appendix:scalar_charge}.

\paragraph{Quadrupole deviations from Kerr ($n_r = -2$).}
Deviations from the Kerr quadrupole moment enter the waveform as a weak-field correction with power-law index $n_r=-2$ (a 2PN effect)~\cite{1995PhRvD..52..821K, 2008PhRvD..78d4013P}. As a result, they are more tightly constrained for higher-mass primaries, which accumulate more cycles in the strong-field regime prior to plunge (see Appendix~\ref{sec:AppendixNoHair}). 
Using the notation of~\cite{2007PhRvD..75d2003B} and the energy flux of~\cite{2006PhRvD..73f4037G}, the forecasted constraint $\sigma_A \simeq 10^{-5}$ for the $m_1=10^6\,M_\odot$ system translates to $\sigma_{\delta Q}\simeq 5\times10^{-4}$, given $\delta Q \simeq 50\,A$ (i.e., the deviation from the Kerr quadrupole moment is $\delta Q$). This is tighter than the bounds shown in Fig.~1 of Ref.~\cite{2007PhRvD..75d2003B}---which employed a kludge model, one year of data, $m_2=100\,M_\odot$, and $\snr=100$---and is consistent with the ranges reported in Fig.~13 of~Ref.~\cite{2017PhRvD..95j3012B} and Fig.~1 of Ref.\cite{2021PhRvD.104f4008Z}.

For comparison, the ringdown analysis of GW250114~\cite{2025PhRvL.135k1403A} yields $|\delta Q|\lesssim 0.3$, corresponding to $|A|_{\rm GW250114}\lesssim 6\times10^{-3}$. Likewise, GW241011~\cite{2025ApJ...993L..21A} provides a precise measurement of the primary spin, implying an upper bound on the spin-induced quadrupole deviation $|\delta Q|\lesssim 0.17$, or $|A|_{\rm GW241011}\lesssim 3\times10^{-3}$. The EMRI constraint from the $(m_1,m_2)=(10^5,10)\,M_\odot$ system improves on these ringdown bounds by $\sim 2$ orders of magnitude, and the higher-mass EMRI by roughly an additional order of magnitude.

We caution the reader against interpreting these comparisons as indications that LISA performs better than ground-based detectors, or vice versa. These constraints probe complementary regions of the binary parameter space and rely on different portions of the gravitational-wave signal.

\subsubsection*{Beyond-GR requirement}
For environmental effects (large positive $n_r$), the loosest constraints are set by the $m_1,m_2=(10^6,10^2)\,M_\odot$ binary; 
for the no-hair test ($n_r = -2$), the loosest constraint comes from the $m_1,m_2=(10^5,10)\,M_\odot$ binary.
An example requirement across all effects is:

\emph{The LISA performance shall ensure that the measurement precision of the inspiral rate deviation $A$ at each slope $n_r$, as defined in Eq.~\eqref{eq:modifiedtorque}, does not increase by more than 40\% relative to the baseline evaluation for the two reference binary systems in Fig.~\ref{fig:precision_nr} for $T=4.5$ years.
} 
\noindent Validation: run the pipeline for the two reference sources with the updated PSD and verify that $\sigma_A(d) / \sigma_A^{\rm ref} \leq 1.4$. These two sources are taken as representative of the most favorable configurations for probing deviations from vacuum GR, for the reasons discussed above. In particular, the low-mass EMRI probes the early inspiral, corresponding to more positive values of $n_r$, while the higher-mass system probes the region closer to the horizon. Therefore, if even these benchmark sources fail to satisfy the requirement, it is unlikely that any other system would enable this science investigation. This is justified by the fact that the two reference systems bracket the relevant parameter space: e.g. if one expects weaker constraints for a $(m_1,\, m_2) = (10^6,\, 10)$ EMRI (since the frequency evolution is slower), this behavior should already be captured by the two reference sources, ensuring that the requirement is conservative. 
Moreover, since beyond-vacuum GR science requires the full 4.5-year mission, any reduction in mission duration introduces an additional risk to this investigation, as quantified by the mission-duration requirement in Section~\ref{subsec:mission_duration}.

\section{Conclusions}
\begin{figure*}
    \includegraphics[]{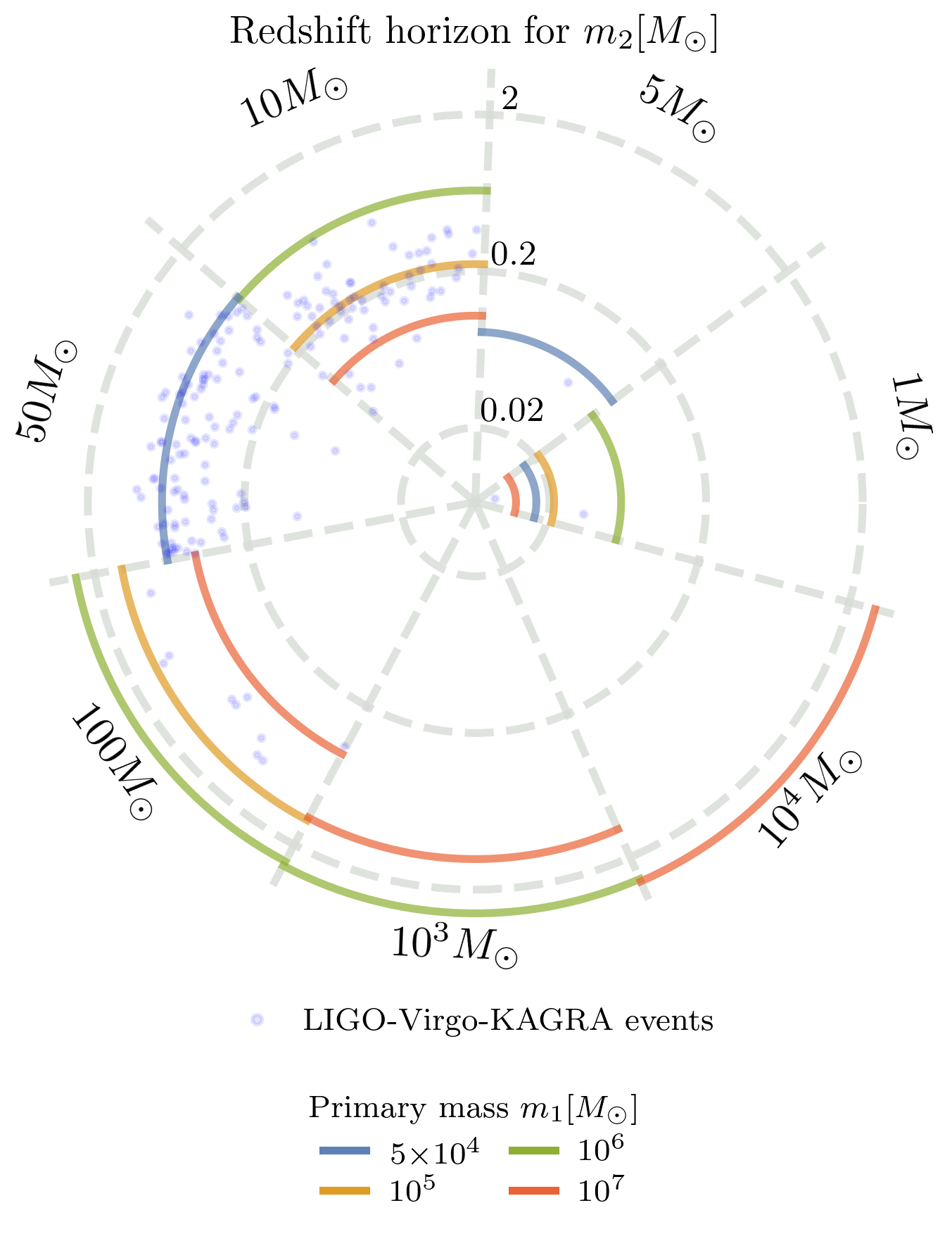}
    \hspace{0.1cm}
    \includegraphics[]{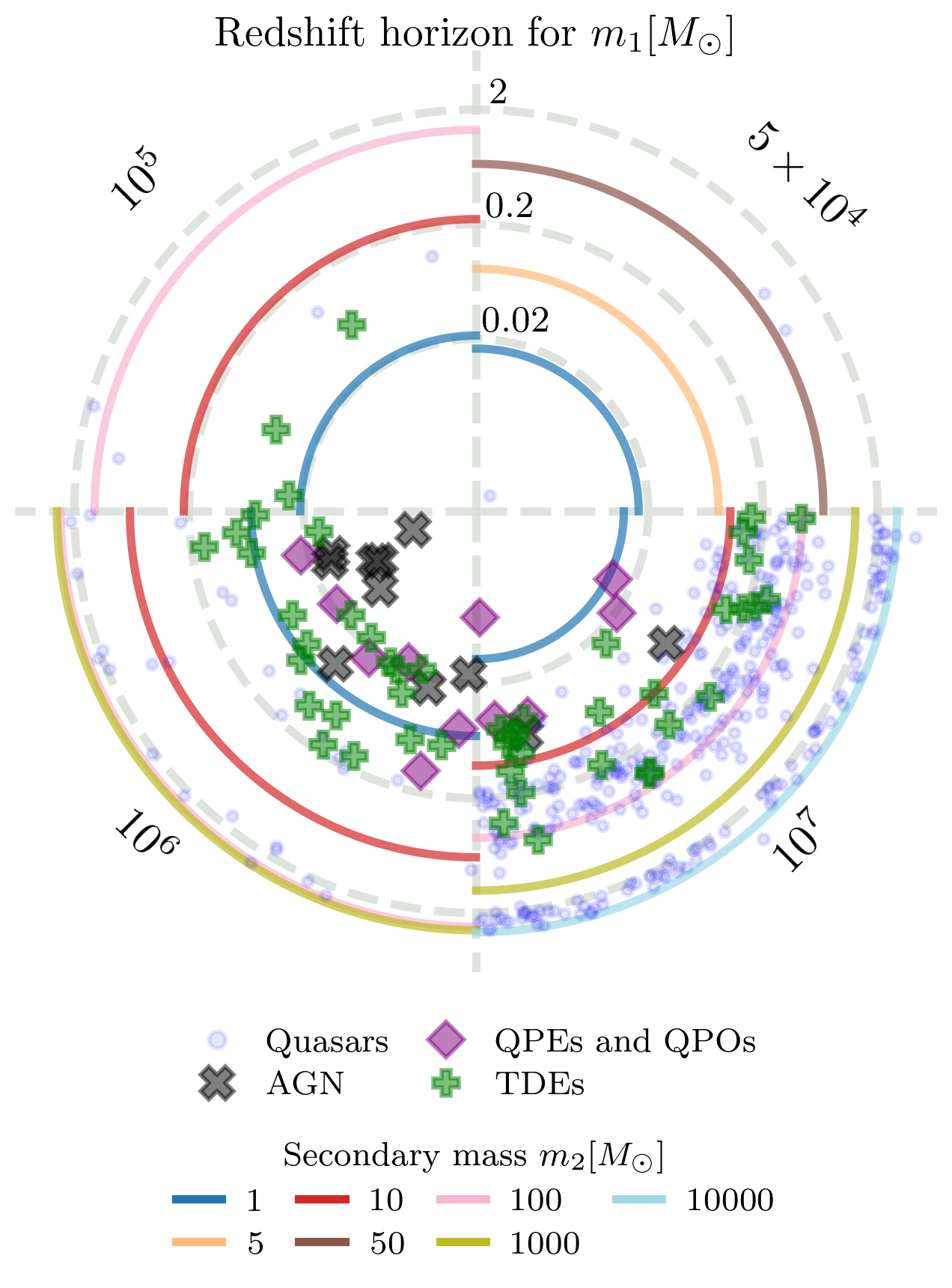}
    \caption{
    Science horizon for EMRI/IMRI sources with prograde spin $a=+0.99$ and time to plunge of 3 months for different source-frame primary masses $\mone$ and secondary masses $\mtwo$ for a detection threshold of $\snr = 30$.
    Left: probes of compact objects around galactic centers compared to the binned LIGO-Virgo-KAGRA mass distribution.
    Right: probes of massive black holes in the local universe compared to the binned electromagnetic observations 
    of Quasars (blue dots), Active Galactic Nuclei (AGN) (black cross), Quasi Periodic Eruptions and Oscillations (QPE and QPO) (violet diamond), and Tidal Disruption Events (TDEs) (green plus).
    }
    \label{fig:horizon_polar}
\end{figure*}
This work develops a pipeline linking LISA instrument performance to the 
EMRI/IMRI science objectives and derives quantitative performance thresholds 
for detection, parameter estimation, and tests of beyond-vacuum GR. 
Figure~\ref{fig:horizon_polar} summarises the science reach under the 
conservative three-month baseline.
The left panel shows the redshift horizon 
for probing stellar-remnant and intermediate-mass companions around galactic 
nuclei, overlaid with the LIGO-Virgo-KAGRA mass distribution.
The right 
panel compares the horizon for massive black hole characterization with 
existing electromagnetic samples, highlighting 
LISA's unique coverage of the low-mass end currently inaccessible to 
electromagnetic surveys.
The main findings of this work are:

\begin{itemize}

  \item \textbf{Detection:} With the baseline PSD, the reference EMRI/IMRI 
  systems can be detected at SNR~$\geq 30$ with three months of data out to 
  $z \sim 0.01$--$0.1$ for EMRIs and $z \sim 2$ for heavy IMRIs ($m_1=10^6 M_\odot$). 
  The detection horizon peaks at 
  $m_1 \sim 10^6\,M_\odot$ and is most prone to sensitivity degradation 
  for light secondaries ($m_2 = 1\,M_\odot$) on retrograde orbits. 

  \item \textbf{Parameter estimation:} At median SNR~$= 30$ and 
  prograde orbits, source-frame masses are measured at 
  $\mathcal{O}(10^{-2})$ precision, dominated by the luminosity distance 
  uncertainty; detector-frame masses reach $\sigma/m \sim 10^{-5}$--$10^{-2}$.
  Spin is constrained to $\sigma_a/a \sim 10^{-6}$--$10^{-3}$, enabling 
  precision tests of the Thorne limit. Sky localization ranges from 
  $30$--$500\,\mathrm{deg}^2$ and luminosity distance precision is 
  $11$--$32\%$. 
  Initial eccentricity relative precision stays below $20\%$ for mass ratios $m_2/m_1=10^{-3}$. 
  The first sources 
  impacted by sensitivity degradation are IMRIs, due to their shorter 
  in-band duration.

  \item \textbf{Mission duration:} Extending observations from 3 months to 
  4.5 years increases the detection horizon by a factor of $\sim 4$ for 
  $m_2 = 1\,M_\odot$ systems and improves sky localization by one to two 
  orders of magnitude, reaching precision 
  below ten squared degrees. Mass 
  precision improves by more than an order of magnitude for low-mass secondaries 
  and spin precision by up to two orders of magnitude for extreme mass ratios. These 
  gains are compounding: longer baselines simultaneously increase SNR, 
  break parameter degeneracies, and amplify the Doppler modulation that 
  encodes sky position.

  \item \textbf{Beyond-vacuum GR and astrophysical environments:} With the 
  full 4.5-year mission, LISA can constrain the fractional inspiral-rate 
  deviation $A$ to $\sigma_A \sim 10^{-5}$ for scalar charge 
  ($n_r = 1$, improving on GW230529 by roughly one order of magnitude) and 
  $\sigma_{\delta {Q}} \sim 5 \times 10^{-4}$ for quadrupole deviations from 
  Kerr ($n_r = -2$). Environmental effects from accretion disks ($n_r = 8$) 
  and dark matter overdensities ($n_r = 5.5$) are best constrained by 
  lower-mass primaries, which probe larger orbital separations, where these 
  effects are enhanced. Overall, LISA can probe a wide range of realistic configurations for these astrophysical environments. 
%   Some of these tests might be compromised for shorter mission durations.

\end{itemize}

More specifically for the science investigations, the main implications of these results are as follows.
For SI~3.1, the EMRI horizon across $m_1\sim 10^5-10^7\,M_\odot$ for a secondary object of $m_2\sim 10$ is between redshifts $z\sim 0.1-0.5$ (right panel of Fig.~\ref{fig:horizon_polar}), and the massive black hole detector-frame mass and dimensionless spin measurements are below $1\%$ and $0.1\%$, respectively (Fig.~\ref{fig:precision_m1_a}).
For SI~3.2, the IMRI horizon in the $m_2\sim10^3$--$10^4\,M_\odot$ range is $z\sim 1-3$ (left panel of Fig.~\ref{fig:horizon_polar}).
For SI~5.2--5.5 (including SI~5.3), the projected bounds on inspiral-rate deviations for representative slopes $n_r=8,\,5.5,\,1,\,-2$ show that LISA can test Kerr consistency and beyond-GR dissipation channels associated with new fields or additional GW emission mechanisms, with stronger constraints than current ground-based constraints for the full 4.5-year observing phase.

Example performance criteria emerge from this analysis: (i) SNR shall not 
decrease by more than 30\% relative to baseline ($d \leq 2$) for any 
$\{m_1, a\}$ configuration with $m_2 = 1\,M_\odot$; (ii) parameter 
uncertainties shall not increase by more than 40\% relative to baseline 
($d \leq 2$) for any source in the reference population; 
(iii) mission duration and duty cycle shall not decrease below the 4.5-year 
nominal phase; and (iv) the $\sigma_A$ precision shall not exceed the envelope of Fig.~\ref{fig:precision_nr} for the slopes considered.

Future work should extend the parameter-space coverage to IMRIs with 
$m_1 < 5 \times 10^4\,M_\odot$, incorporate secondary spin (which will 
likely worsen IMRI mass measurements by an order of magnitude), resonances \cite{2021PhRvD.103l4032S}, orbital 
inclination, and astrophysically motivated distributions of eccentricity 
and time-to-plunge. 
Beyond EMRIs and IMRIs, this study should be extended to similar sources like extreme-mass ratio bursts \cite{2020MNRAS.498L..61H,2013ASPC..467..185B,2025PhRvD.111h3010C} and extremely large mass ratio inspirals \cite{2019PhRvD..99l3025A}.

Regarding beyond vacuum effects, we identify three key developments needed.
Firstly, an extensive exploration of IMRIs potential to probe beyond-vacuum GR.
Secondly, a parametrized post-Einsteinian framework~\cite{2009PhRvD..80l2003Y} for IMRIs/EMRIs in generic orbits is missing and will be crucial for allowing tests of fundamental physics.
Thirdly, a more comprehensive mapping between the parametrized 
inspiral deviation and specific physical models is needed \cite{Gliorio:2026yvh}, as theory-specific 
mappings (e.g.\ scalar charges in scalar Gauss-Bonnet 
theories~\cite{2026PhRvD.113b3036S})
can produce different 
constraints from the agnostic power-law analysis presented 
here~\cite{2017PhRvD..96h4039C}.

Taken together, this work establishes a population-independent framework for exploring the scientific potential of EMRIs and IMRIs and how these are affected by the LISA instrument performance. 
The pipeline and illustrative performance criteria derived here go beyond a single snapshot because they are tied to observable metrics rather than assumed event rates.
They remain valid across a wide range of astrophysical uncertainties and can be updated as the waveform and LISA instrument models mature.
% EMRIs and IMRIs as gravitational-wave sources simultaneously probe the mass and spin distributions of massive black holes across cosmic time, test the Kerr hypothesis at the level of parts per thousand, and constrain astrophysical environments. 
As LISA advances toward launch, the quantitative thresholds established here provide a direct bridge between instrument design decisions and the depth of the science the mission is capable of delivering.

\section*{Data Availability}\label{sec:data_availability}

This work comes with publicly available \href{https://github.com/lorenzsp/ResponseRequirements}{software} to reproduce the results presented.

\begin{acknowledgments}
L.~S.\ thanks the LISA Science Team and the Performance \& Operation Team at ESTEC for their valuable input. 
L.~S.\ is grateful to Gijs Nelemans for providing the computational resources that enabled the large-scale simulations and data analysis presented in this work. 
L.~S.\ also thanks Matilde Signorini for her assistance in incorporating quasar observations. 
L.~S.\ is supported by the European Space Agency Research Fellowship Programme.
This work used the Dutch national e-infrastructure with the support of 
the SURF Cooperative using grant no. EINF-10027.
This research has made use of data or software obtained from the Gravitational Wave Open Science Center (gwosc.org), a service of the LIGO Scientific Collaboration, the Virgo Collaboration, and KAGRA. 
O. Burke acknowledges financial support from the Grant UKRI972 awarded via the UK Space Agency.
A.S. gratefully acknowledges support by the German Space Agency (DLR) with funding of the Bundesministerium für Wirtschaft und Klimaschutz with a decision of the Deutsche Bundestag (Project Ref. No. FKZ 50
OQ 2301).
C.E.A.C-B. is supported by UKSA grant UKRI971.
The authors acknowledge the use of large language models for proofreading and polishing the manuscript. 
All text was carefully reviewed and edited to ensure accuracy.
This work makes use of the Black Hole Perturbation Toolkit.
\end{acknowledgments}

\appendix
\section{Electromagnetic Observations}
\label{app:em_observations}
This appendix describes the electromagnetic observations used in this work to contextualize the EMRI/IMRI parameter space accessible to LISA. 
We compile measurements from multiple observational campaigns spanning different mass regimes and wavelengths.
We stress that all the electromagnetic observations are used only as representative and are not intended to provide precise measurements of masses and redshifts.

We utilize the Sloan Digital Sky Survey Data Release 16 Quasar catalog (SDSS DR16Q) for massive black hole mass measurements at cosmological distances. 
The catalog provides black hole masses derived from virial mass estimators using broad emission lines and single-epoch measurements~\cite{2022ApJS..263...42W}.
The SDSS DR16Q catalog reports black hole masses as $\log_{10}(M_{\rm BH}/M_\odot)$ with associated uncertainties $\sigma[\log_{10}M]$, typically of order 0.3-0.5 dex.
We do not use the full catalog, but we consider only observations with masses $\log_{10}(M_{\rm BH}/M_\odot) < 7.05$ relevant for LISA, and
 exclude poorly constrained measurements $\sigma[M]/M > 0.5$.
After these cuts, the sample contains $\mathcal{O}(10^4)$ quasars with black hole masses spanning $10^5 M_\odot \lesssim M_{\rm BH} \lesssim 10^{7.05} M_\odot$ and redshifts $0.01 \lesssim z \lesssim 5$.
Additionally, we include 11 nearby active galactic nuclei from Table~III of Ref.~\cite{2025arXiv250103252L} obtained with black hole mass measurements from X-ray reverberation mapping~\cite{2022ApJS..261....2K, 2009A&A...497..635C}. 

We include 11 sources exhibiting quasi-periodic eruptions or oscillations from Ref.~\cite{2024MNRAS.532.2143K} based on Refs.~\cite{2019Natur.573..381M,2021Natur.592..704A,2024A&A...684A..64A,2020sea..confE..40G,2008arXiv0807.1899G,2013ApJ...776L..10L,2019Sci...363..531P,2024SciA...10J8898P,2024NatAs...8..347G}. 
These X-ray transient phenomena are hypothesized to arise from orbiting bodies near the innermost stable circular orbit of massive black holes, making them potential electromagnetic counterparts to EMRI systems.
The QPE/QPO sample spans $M_{\rm BH} \sim 10^5$--$10^8 M_\odot$ at redshifts $z \sim 0.005$--$0.13$. 
The masses are derived from various methods including stellar velocity dispersion measurements, X-ray spectral modeling, and optical/UV disk fitting. 
Systematic uncertainties on these masses can be substantial ($\sim 0.5$--$1$ dex, where 1 dex corresponds to a factor of 10 in the linear quantity)~\cite{2024MNRAS.532.2143K}, reflecting the challenges in measuring black hole masses in active systems.

We include the observations of black holes and redshift from Tidal Disruption Events
presented in Ref.~\cite{2024MNRAS.527.2452M}.
Black hole masses are inferred from the late-time optical/UV \emph{plateau} luminosity of TDE
light curves, which arises from an accretion disc forming after stellar disruption.

\section{Mapping of beyond-vacuum General Relativity effects\label{appendix:beyond_GR}}

Here, we describe how to map the parametrized power-law model for deviations from the vacuum General Relativity driven EMRI evolution in Eq.~\eqref{eq:modifiedtorque} to physical quantities of interest (and vice-versa).
Each physical effect is represented by a typical index slope ($n_r$).
In the following, we will be using rescaled angular momentum fluxes as done in Eq.~\eqref{eq:modifiedtorque}.
%%%%%%%%%%%%%%%%%%%%%%%%%%%%%%%%%%%%%%%%%%%%%%%%%%%%%%%%%%%%%%%%
\subsection{Disk Migration Torques ($n_r = 8$)}\label{sec:AppendixDisk}
%%%%%%%%%%%%%%%%%%%%%%%%%%%%%%%%%%%%%%%%%%%%%%%%%%%%%%%%%%%%%%%%

The first effect we consider is the interaction of EMRIs with a surrounding gaseous medium forming an accretion disk around the massive black hole. A significant fraction of galaxies ($\gtrsim 10\%$) are accreting at high rates, forming highly luminous active galactic nuclei (AGN)~\cite{2013LRR....16....1A} whose disks can be sufficiently dense to influence the evolution of EMRIs. 

The dominant environmental effect for these systems arises from so-called \textit{migration torques}~\cite{1980ApJ...241..425G, 2002ApJ...565.1257T}, which are exerted by spiral density waves generated by the orbital motion of the compact object. These waves propagate through the disk and excite resonances at specific radial locations, leading to an exchange of angular momentum between the EMRI and the disk. The magnitude of these torques depends sensitively on the density profile and thermodynamic properties of the disk~\cite{ 2024ApJ...968...28T}. The most widely used model for radiatively efficient, optically thick, geometrically thin accretion disks is the $\alpha$-disk model~\cite{1973A&A....24..337S}.  In this model, the shear stresses responsible for angular momentum transport (which are driven by turbulence due to magnetohydrodynamic instabilities) are parametrized by an effective viscosity parameter $\alpha \sim 0.01$--$0.1$.

For a given opacity law and equation of state, which vary across different disk regions, the surface density and vertical scale height of the disk are fully determined by $\alpha$ and by the mass accretion rate. The latter is usually expressed as a fraction of the Eddington accretion rate, $f_{\rm Edd} \sim 0.01 - 0.1$. In the inner regions of the disk, where EMRIs are expected to be observable by LISA, migration torques for circular motion (the interaction with the disk tends to damp eccentricity very effectively) can be written as~\cite{2011PhRvD..84b4032K, 2014PhRvD..89j4059B, 2023PhRvX..13b1035S}
\begin{align}
    \dot{L}_{\rm disk} &=
    A \left( \frac{p}{10 m_1} \right)^{8}
    \frac{32}{5}
    \left( \frac{p}{m_1} \right)^{-7/2} , \\
    A &=
    7 \times 10^{-10}
    \left( \frac{\alpha}{0.1} \right)^{-1}
    \left( \frac{f_{\rm Edd}}{0.1} \right)^{-3}
    \left( \frac{m_1}{10^6 M_\odot} \right) ,
\end{align}
We note that relativistic corrections can enhance the magnitude of these torques by up to an order of magnitude in the strong-field region~\cite{2025arXiv251002433D, 2025PhRvD.112l4068H, 2025PhRvD.112l4012H, 2026arXiv260119123D}, where they will not be perfectly described by a power-law. In addition, stochastic fluctuations in the disk torques~\cite{2019MNRAS.486.2754D, 2025arXiv251210893D}, arising for example from asymmetries in the gas flow or from turbulent processes, can induce systematic shifts in the effective radial power-law index $n_r$ inferred from the inspiral dynamics~\cite{2025PhRvD.111j4079C}.

%%%%%%%%%%%%%%%%%%%%%%%%%%%%%%%%%%%%%%%%%%%%%%%%%%%%%%%%%%
\subsection{Dynamical Friction in Dark Matter  ($n_r = 5.5$)} \label{sec:AppendixDM}
%%%%%%%%%%%%%%%%%%%%%%%%%%%%%%%%%%%%%%%%%%%%%%%%%%%%%%%%%%

Next, we consider the effect of a dark-matter medium on EMRI, which can be dense due to the adiabatic contraction of an initial flat dark matter profile after a low-mass seed black hole forms at its center. Similarly to what happens in disks, as the secondary compact object travels through the environment, it experiences a gravitational drag, also known as \textit{dynamical friction}, arising from the formation of a density wake trailing the perturber. This wake exerts a net force opposite to the direction of motion and leads to a secular loss of orbital energy and angular momentum.

In the simplest treatment, the dynamical-friction force can be written in the Chandrasekhar form~\cite{1943ApJ....97..255C},
\begin{equation}
F_{\rm DF} = \frac{4\pi \rho_{\rm DM}\, m_2^2}{v^2} \,  I(v) 
\end{equation}
where $\rho_{\rm DM}$ is the local DM density, $v$ is the relative speed of the compact object with respect to the medium, and $I(v)$ is a dimensionless function related with the geometry of the environment's mass distribution, which is typically of order $\mathcal{O}(1)$. For quasi-circular motion, this drag torque is conveniently expressed as an effective angular-momentum loss rate per unit mass $m_2$, 
\begin{equation}
\dot{L}_{\rm DF} = 4\pi\, I(v) m_1^2 \rho_{\rm DM} \left(\frac{p}{m_1}  \right)^2 \, , 
\end{equation}
Using the circular-orbit relation $v^2 \sim m_1/p$ (in geometrized units) and rewriting the result in the same factorized form adopted for the disk-migration model, one finds that the leading scaling of the dynamical friction contribution relative to the gravitational-wave angular-momentum flux corresponds to an effective radial power-law index $n_r=5.5$~\cite{2014PhRvD..89j4059B, 2020A&A...644A.147C}
\begin{align}
\dot{L}_{\rm DF} &=  A \left( \frac{p}{10 m_1} \right)^{11/2}
    \frac{32}{5}
    \left( \frac{p}{m_1} \right)^{-7/2} , \\
    A &= 6 \times 10^5 (m_1^2 \rho)  \\ 
    &\approx 6.5 \times 10^{-5} \left(\frac{m_1}{10^6 M_\odot} \right)^2 \left( \frac{\rho_{\rm DM}}{10^{18} \, M_\odot / \rm pc^3} \right) \, ,
\end{align}
where we implicitly assumed $I(v) \sim \mathcal{O}(1)$. 

As in disks, relativistic treatments of dynamical friction enhance its magnitude by orders of magnitude in the strong-field region~\cite{2022PhRvL.129x1103C, 2025PhRvL.135u1401V, 2022PhRvD.105f1501C, 2024PhRvL.133l1404D, 2025PhRvL.134u1403D, 2023PhRvD.108h4019B}.

\subsection{Secondary scalar charge ($n_r = 1$)} \label{appendix:scalar_charge}

In addition to astrophysical environmental effects, EMRIs can also be affected by modifications to General Relativity and by the presence of new fundamental fields~\cite{2024PhRvD.109j4079D, 2023PhRvL.131e1401B,2022PhRvD.106d4029B,2022NatAs...6..464M,2020PhRvL.125n1101M,2024PhRvD.109f4022S,2026PhRvD.113b3036S,2024PhRvD.109f4022S,Gliorio:2026yvh}.
By following the approach developed in Ref.~\cite{2020PhRvL.125n1101M}, we consider a broad class of theories with a massless (shift-symmetric) scalar field non-minimally coupled to gravity, described by the action
\begin{align}
\mathcal{S}
&=
\int d^4x \sqrt{-g}
\left[
\frac{R}{16\pi G}
-
\frac{1}{2}\partial_\mu \phi \partial^\mu \phi
\right]
+
\alpha_c\, \mathcal{S}_c\!\left[ \bm{g}, \, \phi \right]
\nonumber
\\
&\quad+
\mathcal{S}_m[\bm{g},\phi, \Psi],
\label{eq:scalar_action}
\end{align}
where \(R\) is the Ricci scalar, $\bm{g}$ is the metric tensor, \(\phi\) is a massless scalar field, \(\alpha_c\) is a theory-dependent coupling constant with dimensions $[\mathrm{mass}]^{n}$, where $n > 1$, \(\mathcal{S}_c\left[ \bm{g}, \, \phi \right]\) encodes scalar--gravity interactions, and \(\mathcal{S}_m\) denotes the matter action of the matter fields $\Psi$. We assume that $\mathcal{S}_c$ is invariant under the shift-symmetry $\phi \rightarrow \phi \,+ $ constant, and that the solutions of the field equations derived from Eq.~\eqref{eq:scalar_action} are continuously connected to GR solutions for $\alpha_c \rightarrow 0$.

In this class of theories, black holes can acquire a dimensionless scalar charge \(d\) $\sim \alpha_c/\textnormal{mass}^n$ (not to be confused with the degradation factor of the LISA sensitivity).
For asymmetric binaries, at leading (adiabatic) order, the scalar charge of the primary black hole, if any~\cite{1970CMaPh..19..276C,1995PhRvD..51.6608B,1972CMaPh..25..167H,2023PhRvD.108f4058C,2012PhRvL.108h1103S,2013PhRvL.110x1104H}, is then suppressed, and the background metric can be adequately described by the Kerr metric. The deviations from GR are instead encoded in the scalar charge of the secondary, which sources scalar radiation and modifies the inspiral dynamics. At adiabatic order, the dissipative gravitational and scalar fluxes are fully decoupled, so that the total energy loss for circular orbits is given by
\begin{align}
&\dot{E} = \dot{E}_{\rm GW} +\dot{E}_{scal} \ , \label{eq:relativistic_scalar_flux}\\
&\dot{E}_{scal} = \sum^\infty_{\ell=1}\sum^\ell_{m=-\ell}  \dot{E}^\infty_{\ell m} +\dot{E}^H_{\ell m}\ \nonumber , 
\end{align}
where $\dot{E}_{scal}$, directly proportional to $d^2$, has been computed with the C++ code presented in~\cite{Gliorio:2026yvh} by summing up all the modes $(\ell,m)$ emitted at infinity $\left(\dot{E}^{\infty}_{\ell m}\right)$ and absorbed by the horizon $\left(\dot{E}^{H}_{\ell m}\right)$ necessary to reach $10^{-8}$ accuracy. 
The fluxes have been computed on the same parameter-space grid of FEW (for $e=0$)~\cite{2025PhRvD.112j4023C} and implemented into the~\texttt{KerrCircEqFluxScalar} trajectory, also available at \cite{Gliorio_ModifiedEMRIWaveforms_2026}.

The scalar charge \(d\) can be mapped onto the parametric model considered here. Indeed, the leading scalar contribution corresponds to dipole radiation, which in the weak-field regime enters at \(-1\)PN order relative to the General Relativity quadrupole flux, so that~\cite{2025PhRvD.112j4001C}
\begin{equation}
\dot{E}^{-1\textnormal{PN}}_{scal} = \frac{d^2}{12}\left(\frac{p}{m_1}\right)^{-4} \quad \ , \quad \dot{L}^{-1\textnormal{PN}}_{scal} = \frac{d^2}{12}\left(\frac{p}{m_1}\right)^{-5/2} \ . 
\end{equation}
The latter reduces to the power-law expression in Eq.~\eqref{eq:modifiedtorque} for $n_r=1$ and \begin{equation}
A
\simeq
\frac{25}{192}\,d^{2}.
\label{eq:A_d_relation}
\end{equation}

We stress that relativistic corrections in the strong-field regime modify the effective behavior, such that the scalar flux is no longer described by a simple power law. 
By fitting the logarithm of the fluxes for $a=0.99$ as function of the logarithm of the semi-latus rectum $p$
in the range $1.55 \leq p \leq 100$ we obtain $\dot{E}_{scal} \sim p^{1.15}\,  p^{-5}$ and $\dot{L}_{scal} \sim p^{1.15}\,  p^{-7/2}$.

For a given theory, the charge \(d\) can be related to the fundamental coupling constant \(\alpha_c\). For instance, in scalar Gauss--Bonnet gravity, assuming that the secondary is a black hole,
\begin{equation}
d \simeq \frac{\alpha_{\rm GB}}{2\mu^2},
\end{equation}
where \(\mu\) is the mass of the secondary charged object. Observations by ground-based detectors, in particular the event GW230529, have placed a stringent upper bound
\(\sqrt{\alpha_{\rm GB}} \lesssim 1.4~\mathrm{km}\)~\cite{2024arXiv240603568S}. For a representative secondary mass \(\mu \simeq 3.6\,M_\odot\), this translates into 
\begin{equation}
d \lesssim \frac{\alpha_{\rm GB}}{2\mu^2} \simeq 0.035 \ , 
\end{equation}
which implies, using the leading-order relation~\eqref{eq:A_d_relation}, a constraint from GW230529 given by 
\begin{equation}
A_{\rm GW230529} \simeq \frac{25}{192} d^2 \lesssim 1.6 \times 10^{-4} . 
\end{equation}
This value provides a useful benchmark for assessing the detectability of scalar dipole radiation in EMRI systems.
However, this constraint is only approximate, 
given that it uses a leading-order mapping between the scalar charge and the power-law amplitude $A$, 
which does not capture the relativistic corrections to the scalar fluxes and the jacobian transformations between the different parametrizations of deviations from GR. 

A more accurate mapping can be obtained by comparing the results in Fig.~\ref{fig:precision_nr} to the agnostic deviation in phase $\delta \varphi$ adopted in Ref.~\cite{2024arXiv240603568S} which is usually plotted versus post-Newtonian orders, equal to the negative of the slope $n_r$.
\begin{figure}[ht]
    \centering
    \includegraphics[width=0.9\columnwidth]{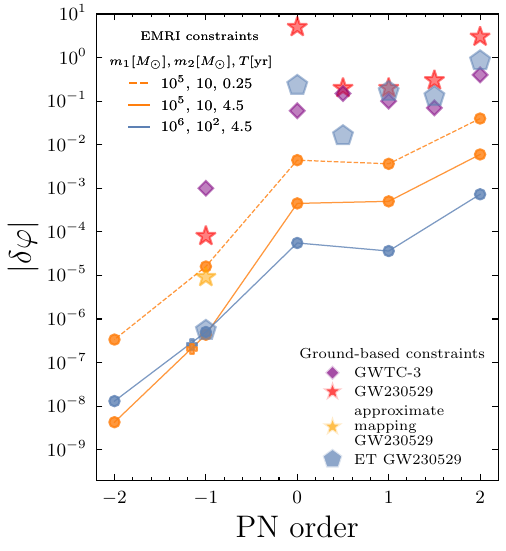}
    \caption{
    Comparison of the constraints on the phase deviation at different post-Newtonian (PN) orders obtained from Fig.~\ref{fig:precision_nr} where we mapped $(A,n_r)\rightarrow (\delta \varphi,\text{PN\, order})$. 
    We show the current constraints obtained from 
    the gravitational wave event GW230529 and using the GW transient catalogs of the third observing run (GWTC-3) \cite{2024arXiv240603568S}. 
    Pentagon markers correspond
to forecasts obtained for a GW230529-like binary observed by the Einstein Telescope (ET) \cite{2024arXiv240607607S}.
    }
    \label{fig:bound_delta_phi}
\end{figure}
The mapping is performed following the steps outlined in Appendix D of Ref.~\cite{2024arXiv240607607S}, or using 
the formalism described in Eq.~(96-97) of Ref.~\cite{2012PhRvD..86b2004C}, Eq.~(21) of Ref.~\cite{2020PhRvD.101j4011N} and Eqs.~(28-29) of Ref.~\cite{2016PhRvD..94h4002Y}.
We show in Fig.~\ref{fig:bound_delta_phi} the constraints on $\delta\varphi$ obtained in Ref.~\cite{2024arXiv240603568S}.
For comparison, we also show the approximate mapping obtained for GW230529 as an orange star, which is smaller than the constraint of Ref.~\cite{2024arXiv240603568S}.
We decide to show in the main text the tightest of the two.
%  to be more conservative in assessing the capability of EMRIs in constraining scalar charges, but we stress that the more accurate mapping is the one obtained by comparing the constraints on $\delta \varphi$ at $-1$PN order.
%%%%%%%%%%%%%%%%%%%%%%%%%%%%%%%%%%%%%%%%%%%%%%%%%%%%%%%%%%%
\subsection{Tests of the No-Hair Theorem ($n_r = -2$)}\label{sec:AppendixNoHair}

Another scientific goal of LISA is to test the nature of compact objects, in particular the Kerr solution. In General Relativity, Kerr black holes obey the no-hair theorem and are fully characterized by their mass $M$ and spin $a$. Equivalently, all their mass and current multipole moments satisfy the relation
\begin{equation}
M_\ell + i S_\ell = M (i M a)^\ell ,
\end{equation}
so that the lowest-order multipoles are given by the mass $M_0=M$, the angular momentum $S_1=J=M^2 a$, and the quadrupole moment
\begin{equation}
Q = M_2 = -M^3 a^2 ,
\end{equation}
with higher multipoles fixed accordingly.

If the primary compact object is not described by the Kerr solution, this relation is violated. We parametrize deviations in the quadrupole moment as
\begin{equation}
Q = -M^3 a^2(1+\delta Q),
\end{equation}
where $\delta Q$ is a dimensionless parameter measuring fractional deviations from Kerr.

The weak-field expansion of the gravitational potential then takes the form
\begin{equation}
\Phi = -\frac{M}{r} - \frac{Q}{2r^3}P_2(\cos\theta) + \cdots ,
\end{equation}
so that the deviation from the Kerr potential is
\begin{equation}
\delta\Phi = -M^3 a^2 \frac{\delta Q}{r^3}.
\end{equation}
This correction enters the equations of motion at order $r^{-3}$, corresponding to an effective 2PN contribution to the inspiral dynamics. In our parametric model for deviations from General Relativity in EMRI inspirals, this implies that quadrupole deviations contribute with exponent $n_r=-2$. 
Using the notation of Ref.~\cite{2007PhRvD..75d2003B} and the energy flux of Ref.~\cite{2006PhRvD..73f4037G}, the associated correction to the angular-momentum flux can be written as
\begin{align}
\dot{L}_{\delta Q}
&=
A
\left(\frac{p}{10m_1}\right)^{-2}
\frac{32}{5}
\left(\frac{p}{m_1}\right)^{-7/2},
\\
A
&\simeq
2\times10^{-2}\,\delta Q \, ,
\end{align}

Recently, the event GW250114~\cite{2025PhRvL.135k1403A} provided the first measurement of an overtone in the post-merger ringdown signal, enabling an independent test of the Kerr nature of the remnant. The observed quasinormal mode frequencies were found to be consistent with the Kerr spectrum at the level
\begin{equation}
\frac{\delta f}{f}\lesssim 30\%.
\end{equation}
In generic parametrized deviations from Kerr, small changes in the quadrupole moment induce linear shifts in the mode frequencies,
\begin{equation}
\frac{\delta f}{f}\simeq C\,\delta Q,
\end{equation}
where $C$ is a dimensionless sensitivity coefficient that depends on the mode and the remnant spin and is typically of $\mathcal{O}(1)$. Assuming this, the GW250114 constraint therefore implies
\begin{equation}
|\delta Q|_{\rm GW250114}\lesssim 0.3.
\end{equation}

Using the mapping above, this translates into a bound on the radiation-reaction amplitude,
\begin{equation}
|A|_{\rm GW250114}\lesssim 6\times10^{-3},
\end{equation}
in our EMRI parametrization.

This estimate is necessarily model dependent, as the precise relation between quasinormal mode spectra and multipole deviations depends on the underlying theory of gravity. Nevertheless, it provides a useful order-of-magnitude connection between existing ringdown tests of the Kerr hypothesis and constraints on inspiral deviations relevant for LISA.

Similarly, the event GW241011~\cite{2025ApJ...993L..21A} yields a precise measurement of the primary black hole spin, setting a non theory-dependent upper bound on the spin-induced quadrupole momentum at 90\% credibility
\begin{equation}
|\delta Q|_{\rm GW241011}\lesssim 0.17.
\end{equation}
which with the mapping above corresponds to 
\begin{equation}
|A|_{\rm GW241011}\lesssim 3 \times 10^{-3}.
\end{equation}
%%%%%%%%%%%%%%%%%%%%%%%%%%%%%%%%%%%%%%%%%%%%%%%%%%

\section{Measurement precision table}\label{app:precision}
We provide in Table~\ref{tab:precision_table} the median measurement precision values of the results presented in Section~\ref{subsubsec:intrinsic_params} for $T=0.25$ years.
% The relative error is undefined for circular orbits

\begin{table*}[ht]
\caption{Median measurement precision of sources with source-frame primary and secondary mass $m_1, m_2$, primary dimensionless spin $a$, initial eccentricity $e_0$, redshift $z$, time to plunge and observation time $T_{\rm pl}=0.25$ years.
}
\label{tab:precision_table}
\centering
\begin{tabular}{ccccccccccccc}
\toprule
$m_1 [M_\odot]$ & $m_2[M_\odot]$ & $a$ & $e_0$ & $z$ & $\Delta \Omega_S [\mathrm{deg}^2]$ & $\sigma_{d_L}/d_L$ & $\sigma_{a}/a$ & $\sigma_{e_0}/e_0$ & $\sigma_{m_{1} }/m_{1}$ & $\sigma_{m_{ 1,\mathrm{det} } }/m_{ 1,\mathrm{det} }$ & $\sigma_{m_{2} }/m_{2}$ & $\sigma_{m_{ 2,\mathrm{det} } }/m_{ 2,\mathrm{det} }$ \\ 
\midrule
5e+04 & 5e+01 & 0.99 & 0.000 & 0.564 & 6.91e+01 & 1.17e-01 & 3.67e-03 & - & 8.39e-02 & 1.79e-02 & 7.14e-02 & 1.15e-02 \\ 
5e+04 & 5e+01 & 0.99 & 0.621 & 0.564 & 9.54e+01 & 1.40e-01 & 5.82e-04 & 2.32e-05 & 6.54e-02 & 3.54e-04 & 6.54e-02 & 8.85e-05 \\ 
5e+04 & 5e+01 & -0.99 & 0.000 & 0.512 & 6.95e+01 & 1.17e-01 & 1.69e-02 & - & 5.95e-02 & 9.20e-03 & 5.42e-02 & 5.84e-03 \\ 
5e+04 & 5e+01 & -0.99 & 0.030 & 0.512 & 6.98e+01 & 1.18e-01 & 1.04e-03 & 1.13e-02 & 5.03e-02 & 3.48e-04 & 5.03e-02 & 1.61e-04 \\ 
1e+05 & 1e+02 & 0.99 & 0.000 & 1.307 & 8.62e+01 & 1.47e-01 & 4.80e-03 & - & 2.67e-01 & 2.89e-02 & 2.19e-01 & 1.85e-02 \\ 
1e+05 & 1e+02 & 0.99 & 0.439 & 1.307 & 8.04e+01 & 1.31e-01 & 4.64e-04 & 6.51e-05 & 1.39e-01 & 3.35e-04 & 1.39e-01 & 7.01e-05 \\ 
1e+05 & 1e+02 & -0.99 & 0.000 & 0.936 & 1.17e+02 & 1.55e-01 & 2.31e-02 & - & 1.36e-01 & 1.21e-02 & 1.26e-01 & 7.50e-03 \\ 
1e+05 & 1e+02 & -0.99 & 0.021 & 0.936 & 8.48e+01 & 1.45e-01 & 6.50e-04 & 3.66e-02 & 1.11e-01 & 1.64e-04 & 1.10e-01 & 1.61e-04 \\ 
1e+06 & 1e+03 & 0.99 & 0.000 & 2.203 & 2.49e+02 & 2.32e-01 & 7.91e-03 & - & 9.80e-01 & 6.81e-02 & 7.65e-01 & 4.25e-02 \\ 
1e+06 & 1e+03 & 0.99 & 0.158 & 2.203 & 2.28e+02 & 2.23e-01 & 4.49e-04 & 3.27e-03 & 4.05e-01 & 6.13e-04 & 4.05e-01 & 2.28e-04 \\ 
1e+06 & 1e+03 & -0.99 & 0.000 & 0.559 & 8.41e+01 & 1.30e-01 & 6.68e-02 & - & 1.24e-01 & 3.22e-02 & 9.11e-02 & 1.83e-02 \\ 
1e+06 & 1e+03 & -0.99 & 0.011 & 0.559 & 8.32e+01 & 1.30e-01 & 8.60e-03 & 7.85e-01 & 6.23e-02 & 3.72e-03 & 6.05e-02 & 1.52e-03 \\ 
1e+07 & 1e+04 & 0.99 & 0.000 & 2.265 & 3.92e+02 & 3.24e-01 & 1.20e-03 & - & 6.84e-01 & 1.84e-02 & 6.31e-01 & 9.85e-03 \\ 
1e+07 & 1e+04 & 0.99 & 0.057 & 2.265 & 3.83e+02 & 3.20e-01 & 2.97e-04 & 1.32e-01 & 5.98e-01 & 1.44e-03 & 5.98e-01 & 1.43e-03 \\ 
1e+07 & 1e+04 & -0.99 & 0.000 & 0.171 & 7.85e+01 & 1.31e-01 & 5.02e-01 & - & 3.33e-01 & 2.30e-01 & 1.65e-01 & 1.10e-01 \\ 
1e+07 & 1e+04 & -0.99 & 0.007 & 0.171 & 8.05e+01 & 1.31e-01 & 3.88e-02 & 7.69e-01 & 3.93e-02 & 1.72e-02 & 2.57e-02 & 7.21e-03 \\ 
5e+04 & 5e+00 & 0.99 & 0.000 & 0.069 & 5.77e+01 & 1.18e-01 & 1.10e-03 & - & 1.41e-02 & 7.40e-03 & 1.15e-02 & 4.71e-03 \\ 
5e+04 & 5e+00 & 0.99 & 0.323 & 0.069 & 5.95e+01 & 1.23e-01 & 1.20e-04 & 4.26e-05 & 8.11e-03 & 9.53e-05 & 8.11e-03 & 1.53e-05 \\ 
5e+04 & 5e+00 & -0.99 & 0.000 & 0.067 & 6.37e+01 & 1.19e-01 & 5.18e-03 & - & 8.74e-03 & 2.60e-03 & 8.09e-03 & 1.55e-03 \\ 
5e+04 & 5e+00 & -0.99 & 0.015 & 0.067 & 6.44e+01 & 1.21e-01 & 4.40e-04 & 3.02e-02 & 7.83e-03 & 1.76e-04 & 7.77e-03 & 7.49e-05 \\ 
1e+05 & 1e+01 & 0.99 & 0.000 & 0.191 & 6.35e+01 & 1.22e-01 & 9.94e-04 & - & 2.62e-02 & 7.50e-03 & 2.38e-02 & 4.75e-03 \\ 
1e+05 & 1e+01 & 0.99 & 0.237 & 0.191 & 6.47e+01 & 1.22e-01 & 7.10e-05 & 8.80e-05 & 2.09e-02 & 6.79e-05 & 2.09e-02 & 1.09e-05 \\ 
1e+05 & 1e+01 & -0.99 & 0.000 & 0.161 & 6.72e+01 & 1.20e-01 & 6.65e-03 & - & 1.85e-02 & 3.25e-03 & 1.78e-02 & 1.89e-03 \\ 
1e+05 & 1e+01 & -0.99 & 0.012 & 0.161 & 6.57e+01 & 1.19e-01 & 2.30e-04 & 6.87e-02 & 1.75e-02 & 9.35e-05 & 1.75e-02 & 1.10e-04 \\ 
1e+06 & 1e+02 & 0.99 & 0.000 & 2.019 & 2.99e+02 & 2.77e-01 & 1.31e-04 & - & 4.59e-01 & 1.89e-03 & 4.59e-01 & 1.03e-03 \\ 
1e+06 & 1e+02 & 0.99 & 0.059 & 2.019 & 2.97e+02 & 2.76e-01 & 3.21e-05 & 1.15e-02 & 4.57e-01 & 1.42e-04 & 4.57e-01 & 1.41e-04 \\ 
1e+06 & 1e+02 & -0.99 & 0.000 & 0.182 & 7.01e+01 & 1.22e-01 & 5.00e-02 & - & 4.73e-02 & 2.29e-02 & 3.07e-02 & 1.10e-02 \\ 
1e+06 & 1e+02 & -0.99 & 0.007 & 0.182 & 6.99e+01 & 1.23e-01 & 4.36e-03 & 8.60e-01 & 2.06e-02 & 1.95e-03 & 2.02e-02 & 8.26e-04 \\ 
1e+07 & 1e+03 & 0.99 & 0.000 & 1.242 & 1.03e+02 & 1.80e-01 & 1.92e-05 & - & 1.82e-01 & 1.60e-03 & 1.81e-01 & 3.14e-04 \\ 
1e+07 & 1e+03 & 0.99 & 0.022 & 1.242 & 1.03e+02 & 1.80e-01 & 3.21e-05 & 2.80e-01 & 1.81e-01 & 4.34e-04 & 1.81e-01 & 5.12e-04 \\ 
1e+07 & 1e+03 & -0.99 & 0.000 & 0.046 & 6.81e+01 & 1.73e-01 & 1.16e+00 & - & 5.74e-01 & 5.20e-01 & 2.16e-01 & 1.98e-01 \\ 
1e+07 & 1e+03 & -0.99 & 0.005 & 0.046 & 6.67e+01 & 1.20e-01 & 4.99e-02 & 6.71e-01 & 2.86e-02 & 2.19e-02 & 1.03e-02 & 5.80e-03 \\ 
1e+05 & 1e+00 & 0.99 & 0.000 & 0.021 & 3.51e+01 & 1.25e-01 & 1.27e-04 & - & 3.36e-03 & 1.24e-03 & 2.95e-03 & 7.53e-04 \\ 
1e+05 & 1e+00 & 0.99 & 0.096 & 0.021 & 3.99e+01 & 1.27e-01 & 8.67e-06 & 2.85e-04 & 2.65e-03 & 1.66e-05 & 2.65e-03 & 1.12e-05 \\ 
1e+05 & 1e+00 & -0.99 & 0.000 & 0.021 & 6.48e+01 & 1.24e-01 & 5.40e-03 & - & 4.40e-03 & 2.47e-03 & 3.12e-03 & 1.19e-03 \\ 
1e+05 & 1e+00 & -0.99 & 0.007 & 0.021 & 6.38e+01 & 1.23e-01 & 7.95e-04 & 3.83e-01 & 2.51e-03 & 3.52e-04 & 2.48e-03 & 1.20e-04 \\ 
1e+06 & 1e+01 & 0.99 & 0.000 & 0.551 & 7.21e+01 & 1.52e-01 & 4.58e-06 & - & 6.96e-02 & 1.75e-04 & 6.95e-02 & 5.90e-05 \\ 
1e+06 & 1e+01 & 0.99 & 0.027 & 0.551 & 7.31e+01 & 1.53e-01 & 4.43e-06 & 1.60e-02 & 6.99e-02 & 2.67e-05 & 6.99e-02 & 4.93e-05 \\ 
1e+06 & 1e+01 & -0.99 & 0.000 & 0.041 & 6.82e+01 & 1.23e-01 & 1.17e-01 & - & 6.15e-02 & 5.23e-02 & 2.52e-02 & 2.00e-02 \\ 
1e+06 & 1e+01 & -0.99 & 0.005 & 0.041 & 6.82e+01 & 1.22e-01 & 9.23e-03 & 5.34e-01 & 8.21e-03 & 4.00e-03 & 5.13e-03 & 1.09e-03 \\ 
1e+07 & 1e+02 & 0.99 & 0.000 & 0.404 & 7.38e+01 & 1.68e-01 & 5.02e-05 & - & 5.78e-02 & 9.95e-04 & 5.76e-02 & 1.00e-03 \\ 
1e+07 & 1e+02 & 0.99 & 0.009 & 0.404 & 7.37e+01 & 1.68e-01 & 1.15e-05 & 5.06e-01 & 5.74e-02 & 1.71e-04 & 5.72e-02 & 3.11e-04 \\ 
1e+07 & 1e+02 & -0.99 & 0.000 & 0.007 & 6.58e+01 & 7.17e-01 & 6.38e+00 & - & 2.89e+00 & 2.85e+00 & 9.15e-01 & 9.07e-01 \\ 
1e+07 & 1e+02 & -0.99 & 0.006 & 0.007 & 6.53e+01 & 1.25e-01 & 1.01e-01 & 3.76e-01 & 4.61e-02 & 4.47e-02 & 9.12e-03 & 8.38e-03 \\ 
1e+06 & 1e+00 & 0.99 & 0.000 & 0.053 & 6.74e+01 & 1.72e-01 & 3.53e-06 & - & 8.89e-03 & 7.80e-05 & 8.91e-03 & 6.74e-05 \\ 
1e+06 & 1e+00 & 0.99 & 0.011 & 0.053 & 6.57e+01 & 1.71e-01 & 1.31e-06 & 1.64e-01 & 8.81e-03 & 2.22e-05 & 8.85e-03 & 7.01e-05 \\ 
1e+06 & 1e+00 & -0.99 & 0.000 & 0.006 & 6.66e+01 & 1.54e-01 & 6.35e-01 & - & 2.87e-01 & 2.83e-01 & 9.10e-02 & 9.02e-02 \\ 
1e+06 & 1e+00 & -0.99 & 0.006 & 0.006 & 6.63e+01 & 1.22e-01 & 1.58e-02 & 7.91e-02 & 7.65e-03 & 6.95e-03 & 1.97e-03 & 1.41e-03 \\ 
1e+07 & 1e+01 & 0.99 & 0.000 & 0.080 & 6.47e+01 & 1.71e-01 & 2.53e-04 & - & 1.53e-02 & 3.33e-03 & 2.50e-02 & 1.02e-02 \\ 
1e+07 & 1e+01 & 0.99 & 0.006 & 0.080 & 6.43e+01 & 1.71e-01 & 3.15e-05 & 2.60e-01 & 1.30e-02 & 3.16e-04 & 1.32e-02 & 1.72e-03 \\ 
1e+07 & 1e+00 & 0.99 & 0.000 & 0.010 & 6.70e+01 & 2.72e-01 & 2.15e-03 & - & 3.12e-02 & 2.66e-02 & 1.28e-01 & 1.24e-01 \\ 
1e+07 & 1e+00 & 0.99 & 0.007 & 0.010 & 6.66e+01 & 1.75e-01 & 4.77e-05 & 4.51e-02 & 1.97e-03 & 5.42e-04 & 6.07e-03 & 4.16e-03 \\ 
5e+04 & 1e+00 & 0.99 & 0.000 & 0.013 & 3.34e+01 & 1.16e-01 & 4.38e-04 & - & 4.93e-03 & 3.61e-03 & 3.58e-03 & 2.26e-03 \\ 
5e+04 & 1e+00 & 0.99 & 0.174 & 0.013 & 3.20e+01 & 1.18e-01 & 1.05e-05 & 6.41e-05 & 1.52e-03 & 1.08e-05 & 1.52e-03 & 6.49e-06 \\ 
5e+04 & 1e+00 & -0.99 & 0.000 & 0.016 & 6.12e+01 & 1.26e-01 & 3.36e-03 & - & 3.03e-03 & 1.59e-03 & 2.35e-03 & 8.74e-04 \\ 
5e+04 & 1e+00 & -0.99 & 0.010 & 0.016 & 6.19e+01 & 1.27e-01 & 4.40e-04 & 1.02e-01 & 2.02e-03 & 1.91e-04 & 2.00e-03 & 6.02e-05 \\ 
\bottomrule
\end{tabular}
\end{table*}

\bibliography{apssamp}% Produces the bibliography via BibTeX.

\end{document}